\documentclass[12pt,preprint]{aastex}

\shorttitle{Nonthermal Filaments
in Historical Supernova Remnants
}
\shortauthors{Bamba et al.}

\begin{document}

\title{A Spatial and Spectral Study of Nonthermal Filaments
in Historical Supernova Remnants:
Observational Results with {\it Chandra}}

\author{
Aya Bamba\altaffilmark{1},
Ryo Yamazaki\altaffilmark{2},
Tatsuo Yoshida\altaffilmark{3},
Toshio Terasawa\altaffilmark{4},
and
Katsuji Koyama\altaffilmark{5}
}
\altaffiltext{1}
{RIKEN (The Institute of Physical and Chemical Research)
2-1, Hirosawa, Wako, Saitama 351-0198, Japan
}
\altaffiltext{2}
{Department of Earth and Space Science,
Graduate School of Science, Osaka University,
Toyonaka, Osaka 560-0043, Japan
}
\altaffiltext{3}
{Faculty of Science,
Ibaraki University, Mito 310-8512, Japan
}
\altaffiltext{4}
{Earth \& Planetary Science, Graduate School of Science,
University of Tokyo,
7-3-1, Hongo, Bunkyo-ku, Tokyo 113-0033, Japan
}
\altaffiltext{5}
{Department of Physics, Graduate School of Science, Kyoto University, 
Sakyo-ku, Kyoto 606-8502, Japan
}

\email{bamba@crab.riken.jp,
ryo@vega.ess.sci.osaka-u.ac.jp,
yoshidat@mx.ibaraki.ac.jp,
terasawa@eps.s.u-tokyo.ac.jp,
koyama@cr.scphys.kyoto-u.ac.jp
}


\begin{abstract}
The outer shells of young supernova remnants (SNRs)
are the most plausible acceleration sites of high-energy electrons
with the diffusive shock acceleration (DSA) mechanism.
We studied spatial and spectral properties close to the shock fronts
in four historical SNRs
(Cas~A, Kepler's remnant, Tycho's remnant, and RCW~86)
with excellent spatial resolution of {\it Chandra}.
In all of the SNRs, hard X-ray emissions were found on the rims of the SNRs,
which concentrate in very narrow regions (so-called "filaments");
apparent scale widths on the upstream side are
below or in the order of the point spread function of {\it Chandra},
while 0.5--40~arcsec (0.01--0.4~pc) on the downstream side
with most reliable distances.
The spectra of these filaments can be fitted with
both thermal and nonthermal (power-law and {\tt SRCUT}) models.
The former requires unrealistic
high temperature ($\ga$2~keV) and low abundances ($\la$1 solar)
for emission from young SNRs
and may be thus unlikely.
The latter reproduces the spectra
with best-fit photon indices of 2.1--3.8,
or roll-off frequencies of (0.1--28)$\times 10^{17}$~Hz,
which reminds us of the synchrotron emission from electrons
accelerated via DSA.
We consider various physical parameters
as functions of the SNR age,
including the previous results on SN~1006 \citep{bamba2003b};
the filament width on the downstream side increases with the SNR age,
and the spectrum becomes softer keeping a nonthermal feature.
It was also found that a function,
that is the roll-off frequency
divided by the square of the scale width on the downstream side,
shows negative correlation with the age,
which might provide us some information on the DSA theory.
\end{abstract}


\section{Introduction}

Ever since the discovery of cosmic rays \citep{hess1912},
the origin and the acceleration mechanism
up to more than $\sim$TeV
have been long-standing problems.
\citet{koyama1995} discovered synchrotron X-rays
from shock fronts of a supernova remnant (SNR), SN~1006,
and suggested that
shocks of SNRs are cosmic ray accelerators.
Several SNRs have recently been categorized 
into synchrotron X-ray emitters
(G347.3$-$0.5: \citet{koyama1997,slane1999},
RCW~86: \citet*{bamba2000,borkowski2001b},
and G266.6$-$1.2: \citet{slane2001}).
Recently,
a significant number of SNRs with synchrotron X-rays have been discovered
by the {\it ASCA} Galactic plane survey
\citep{bamba2001,ueno2003,bamba2003a,yamaguchi2004}.
These discoveries provide good evidence for the cosmic ray acceleration
at the shocked shell of SNRs.
The most plausible process of the cosmic ray acceleration is 
the diffusive shock acceleration (DSA)
\citep[e.g.][]{bell1978,blandford1978,drury1983,%
blandford1987,jones1991,malkov2001},
which can accelerate particles on the shock
into a power-law distribution,
similar to the observed spectrum of cosmic rays
showering on the earth.

Apart from the global success of DSA,
there are still many remaining problems.
We have not yet fully understood detailed but important information,
such as the maximum energy of particles,
the configuration of magnetic fields,
the injection efficiency from thermal plasma to accelerated particles,
and the acceleration and de-acceleration history of particles
around the shock fronts, and so on.
This partly comes from the fact that
there is no information
about the spatial distribution of accelerated electrons of practical objects,
which strongly reflects the above uncertainties.

\citet{bamba2003b} considered
the spatial distribution of emission from accelerated particles
as new information
in order to understand the acceleration on the shock in SNRs;
they observed the synchrotron X-ray emitting shell of SN~1006
with the excellent spatial resolution of the {\it Chandra} X-ray observatory
\citep[see also][]{long2003},
and found that
the emitting regions of synchrotron X-rays are 
incredibly thin ("filaments"),
about 0.01--0.1~pc on the upstream side
and 0.06--0.4~pc on the downstream side,
at 2.18~kpc distance \citep{winkler2003}.
This result constrains the DSA theory and magnetic-field configurations;
both an efficient acceleration scenario with strong magnetic field
parallel to the shock normal 
and an inefficient one with a weak and perpendicular magnetic field
are allowed \citep{yamazaki2004b,berezhko2003}.

In the present work, 
we searched for nonthermal filaments in other historical SNRs
(Cas~A, Kepler's remnant, Tycho's remnant, and RCW~86)
and measured their scale widths and spectral parameters.
It was also investigated
whether or not the age of the SNR is correlated with 
the scale width and/or spectral parameters of filaments.
If it is, such correlations might describe the time evolution 
of these parameters in a single SNR.
Although the characteristics of the SNR should be considered,
those correlations may also  become  clues to understand
the time evolution of the maximum energy of electrons,
the magnetic field configuration,
the injection efficiency, and so on.
This paper is organized as follows.
We summarize the observation details of five SNRs with {\it Chandra}
in \S\ref{sec:obs}.
\S\ref{sec:individual} briefly describes
the results of the observations of the four SNRs.
In \S\ref{sec:discussion}
we discuss the origin of emission (\S\ref{sec:origin}),
our analysis of individual SNRs
(\S\ref{sec:individual2}),
the time evolution of the spatial and spectral parameters of filaments
together with previous analysis of SN~1006,
which have already been analyzed in \citet{bamba2003b}
(\S\ref{sec:evo}).
We also briefly deal with the time evolution of energy densities of
magnetic field and cosmic rays in this section.
A summary of our results is given in \S\ref{sec:sum}.

\section{Observations}
\label{sec:obs}

We used the {\it Chandra} archival data of the ACIS
of five historical SNRs
(Cas~A, Kepler, Tycho, SN~1006, and RCW~86),
as listed in Table~\ref{tab:obs_log}.
Note that this data set includes almost historical shell-type SNRs.
The satellite and the instrument are described
by \citet{weisskopf2002} and \citet{garmire2000}, respectively.   
Data acquisition from the ACIS was made in the Timed-Exposure Faint mode.
The data reductions and analysis were made using the {\it Chandra} 
Interactive Analysis of Observations (CIAO) software version 2.3.
Using the Level 2 processed events provided by the pipeline
processing at the {\it Chandra} X-ray Center,
we selected {\it ASCA} grades 0, 2, 3, 4, and 6, as the X-ray events.
The effective exposure for each observation is also given
in Table~\ref{tab:obs_log}.

\section{Individual Results}
\label{sec:individual}

\subsection{Cas~A (G111.7$-$2.1)}

Although the remnant's precise age is still uncertain,
the most probable record of the explosion
is SN~1680, made by \citet{flamsteed1725},
which we adopt in this paper.
The distance to the remnant was estimated to be $3.4_{-0.1}^{+0.3}$~kpc
by \citet{reed1995}
using the measurement of optical proper motions.
The average radio index of the whole remnant at 1~GHz is
$0.78$ \citep{green2004}.

Cas~A emits hard X-rays up to greater than 10~keV,
which were detected by {\it RXTE} \citep{allen1997},
{\it Beppo}SAX \citep{favata1997},
and OSSE onboard {\it GRO} \citep{the1996}.
\citet{vink2003} has already found and analyzed
one of the nonthermal filamentary structures
with {\it Chandra}.
There are many other filaments thinner than that which they analyzed.
These filaments are considered to be profiles of a thin sheet-like structure,
as suggested by \citet{hester1987};
then, the scale width of these filaments is
the upper-limit width of the real thickness
in the radial direction of the sheets.
Therefore, we consider other filaments in this paper,
which are located much nearer to the aim-point (less than 25 arcsec)
and are much thinner than those analyzed by \citet{vink2003}.

\begin{figure}[hbtp]
\epsscale{0.45}
\plotone{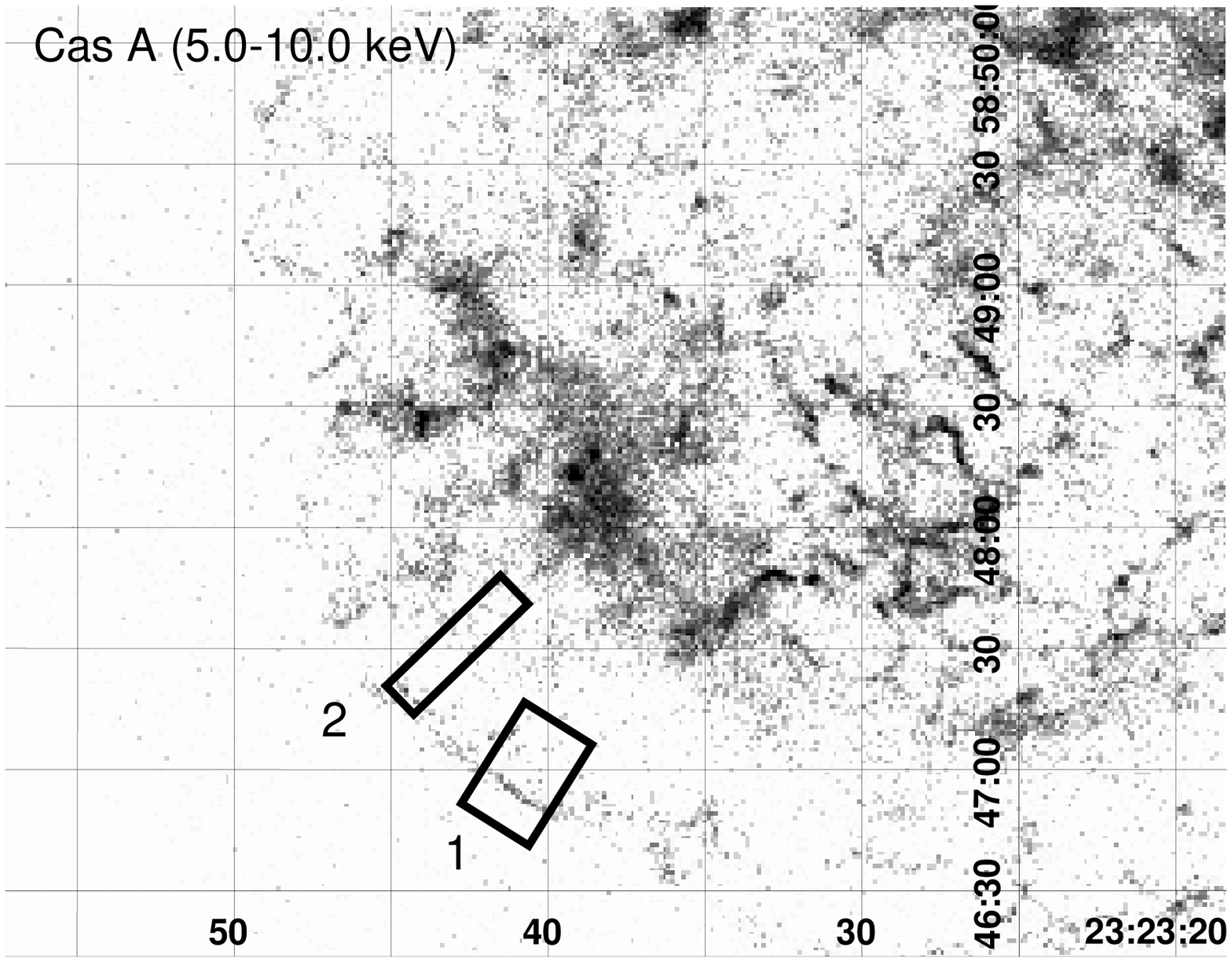}
\plotone{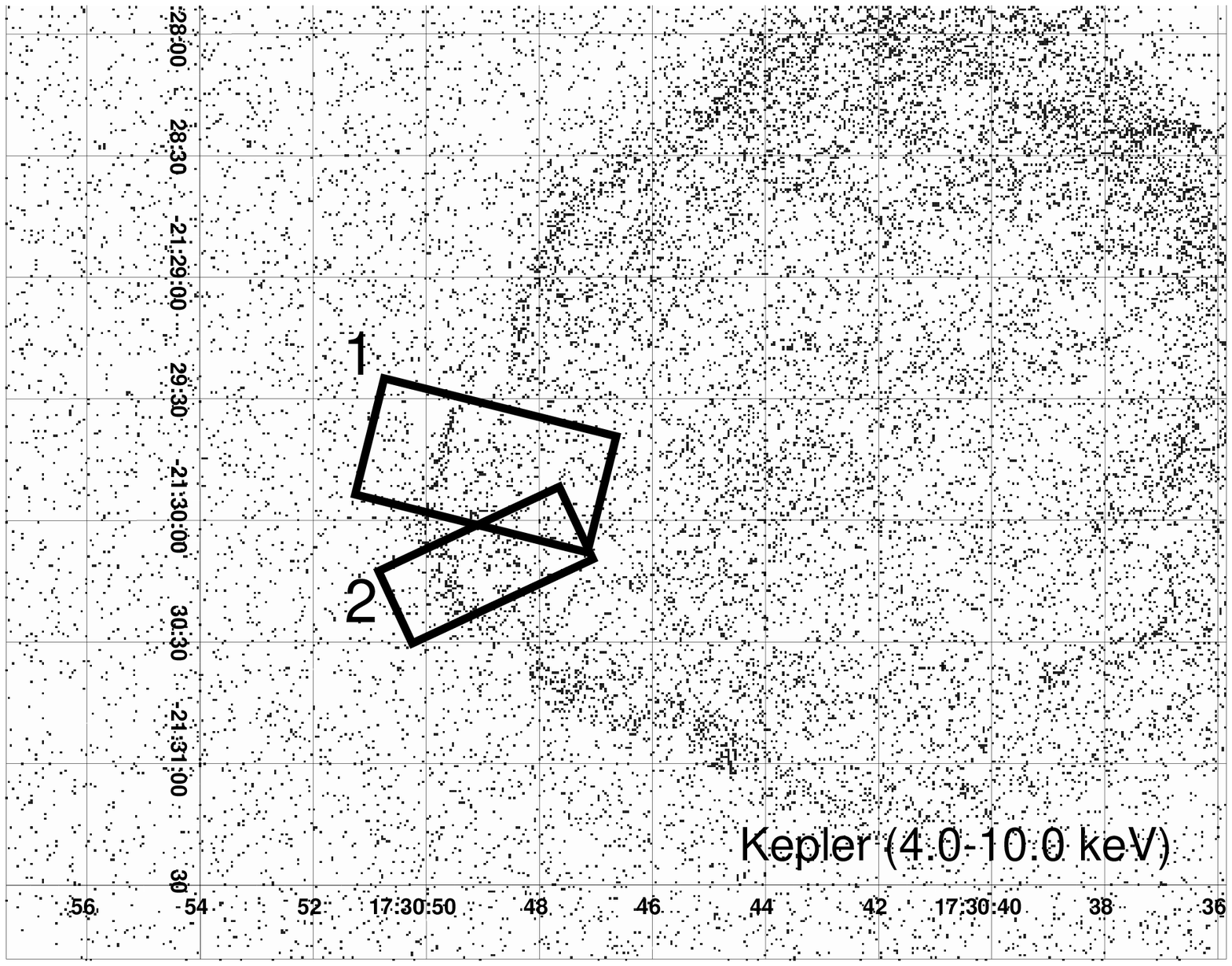}
\plotone{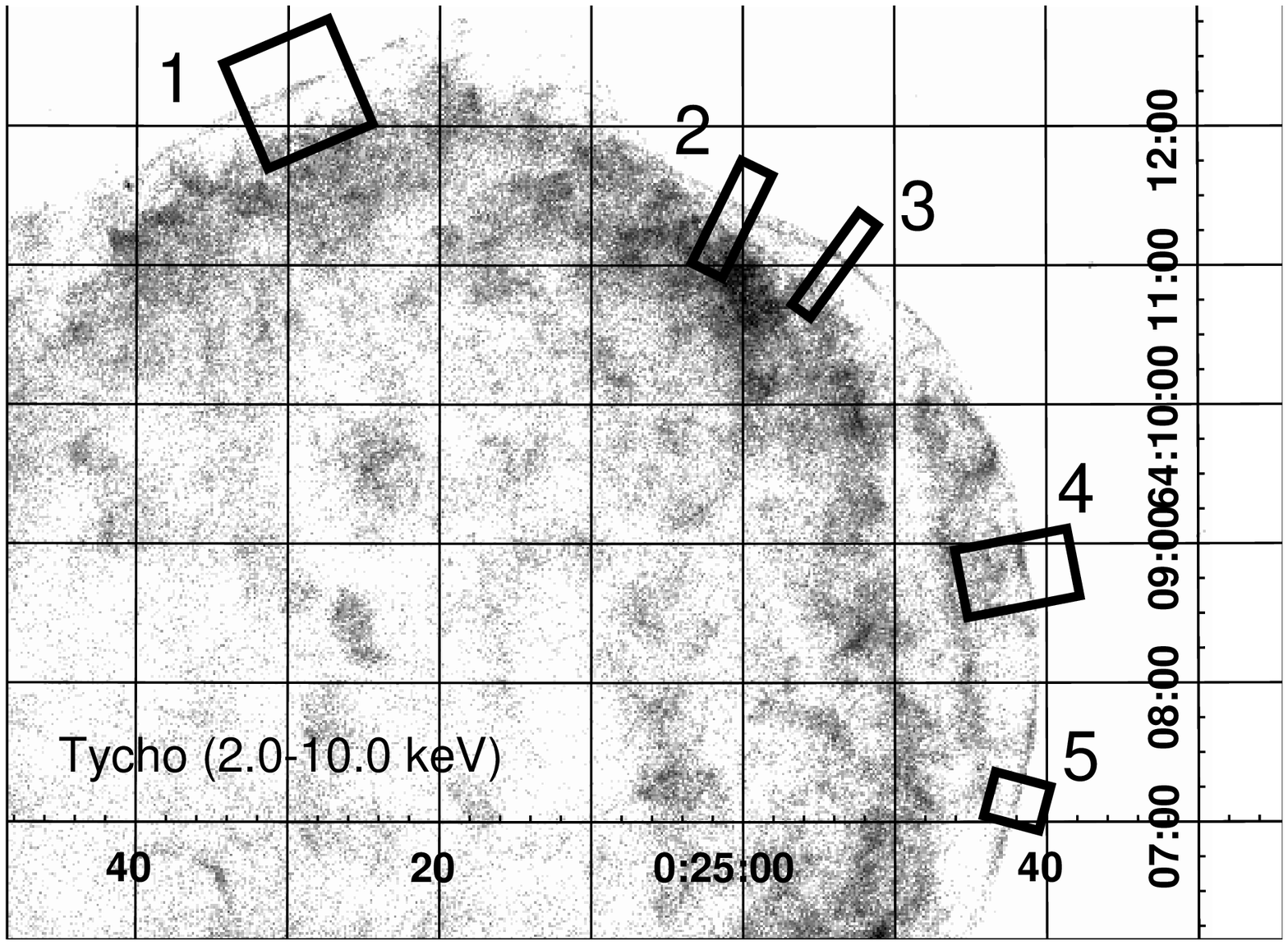}
\plotone{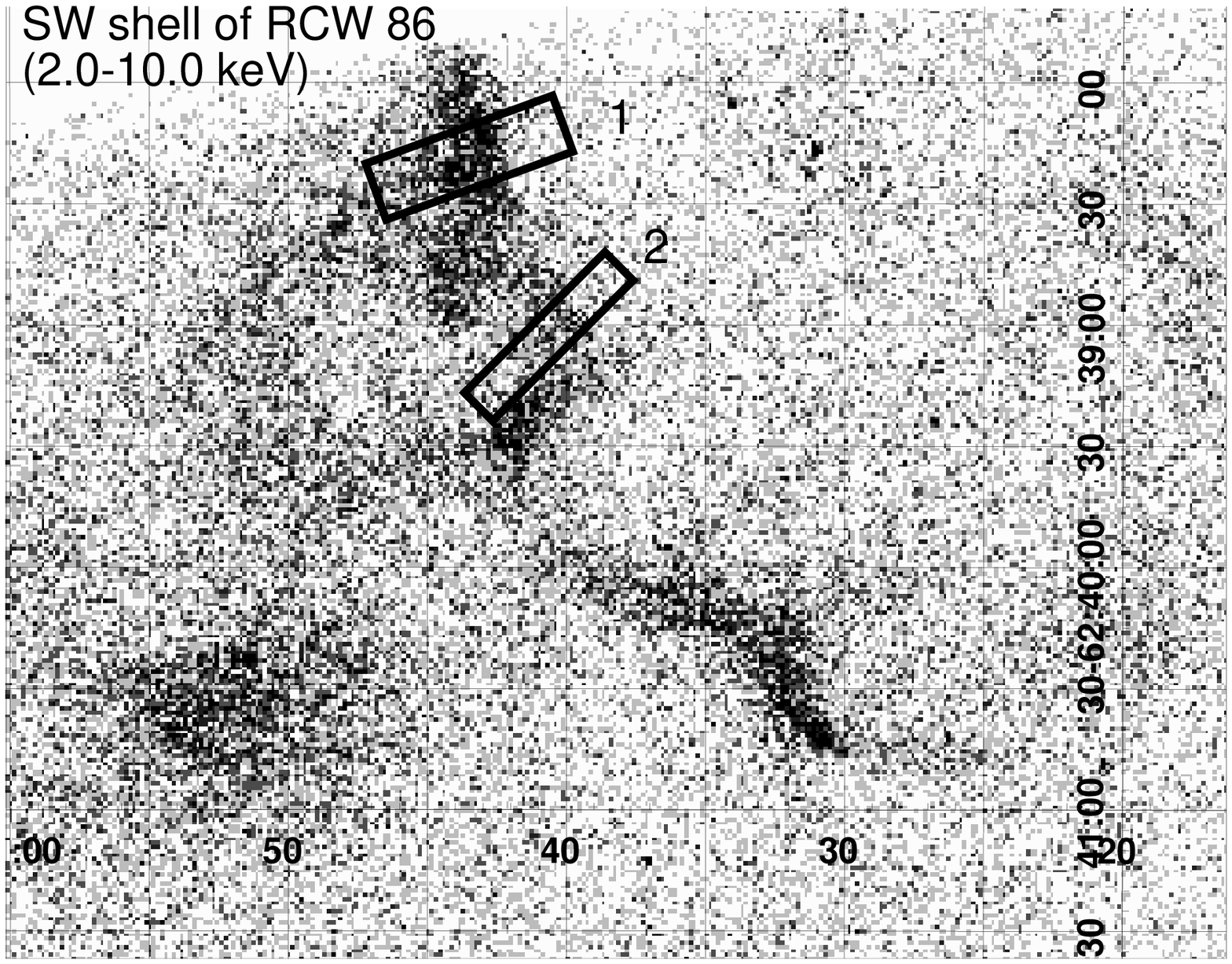}
\caption{Close-up views of the SNRs in the hard X-ray band
(written in each image; see also text) with {\it Chandra}.
Grayscales are in logarithmic and coordinates are in J2000.
The filament regions for spatial and spectral analysis are also shown
with solid rectangles.}
\label{fig:images}
\end{figure}

The upper-left panel of Figure~\ref{fig:images}
shows the south-eastern side of Cas~A in the 5.0--10.0~keV band,
in which there are a lot of filamentary structures.
These filaments are more clumpy than those in SN~1006 \citep{bamba2003b}.
We selected two filaments as shown in the panel,
which are straight enough to be made profiles
and near the aim-point.

\begin{figure}[hbtp]
\epsscale{0.3}
\plotone{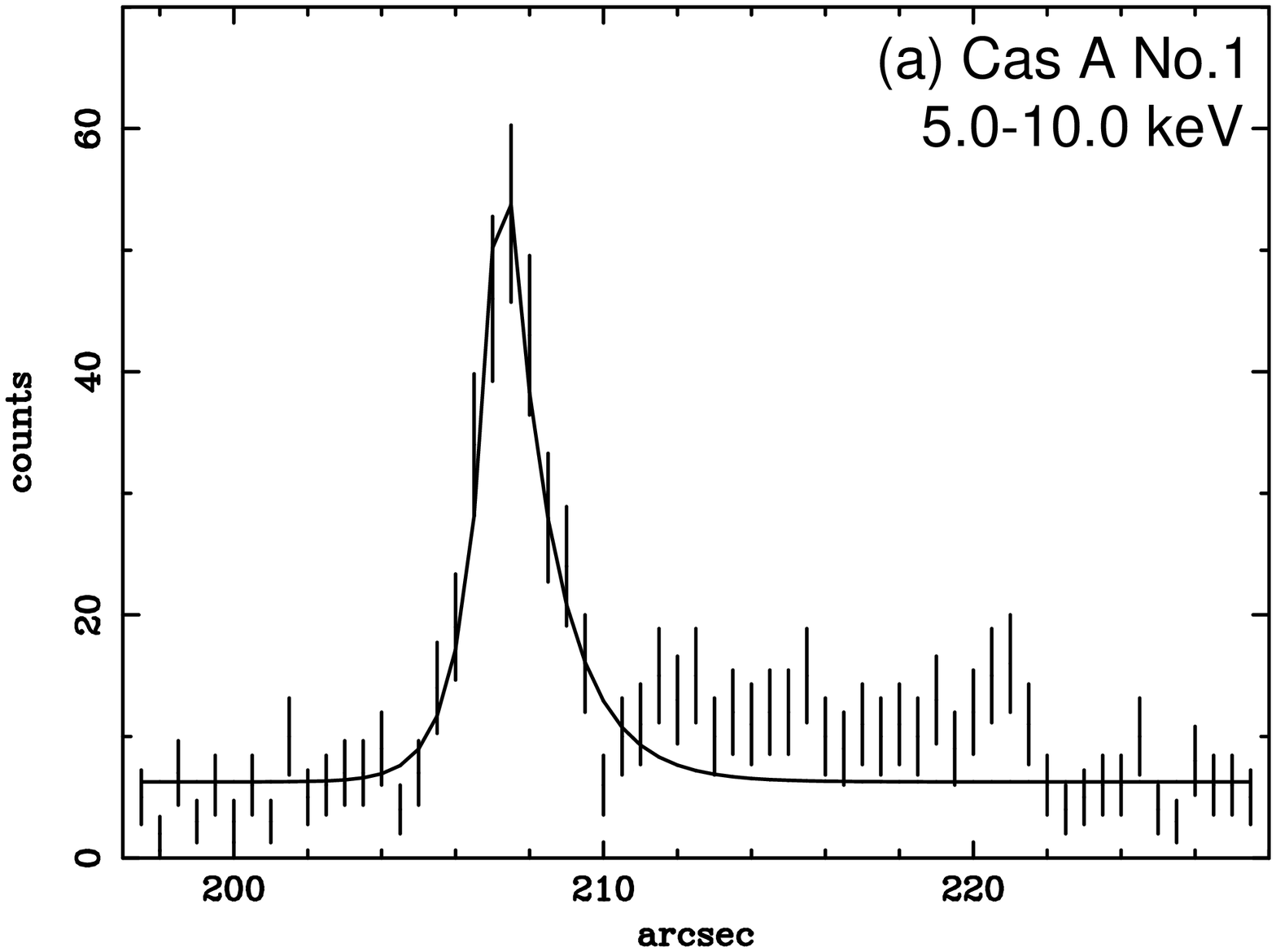}
\plotone{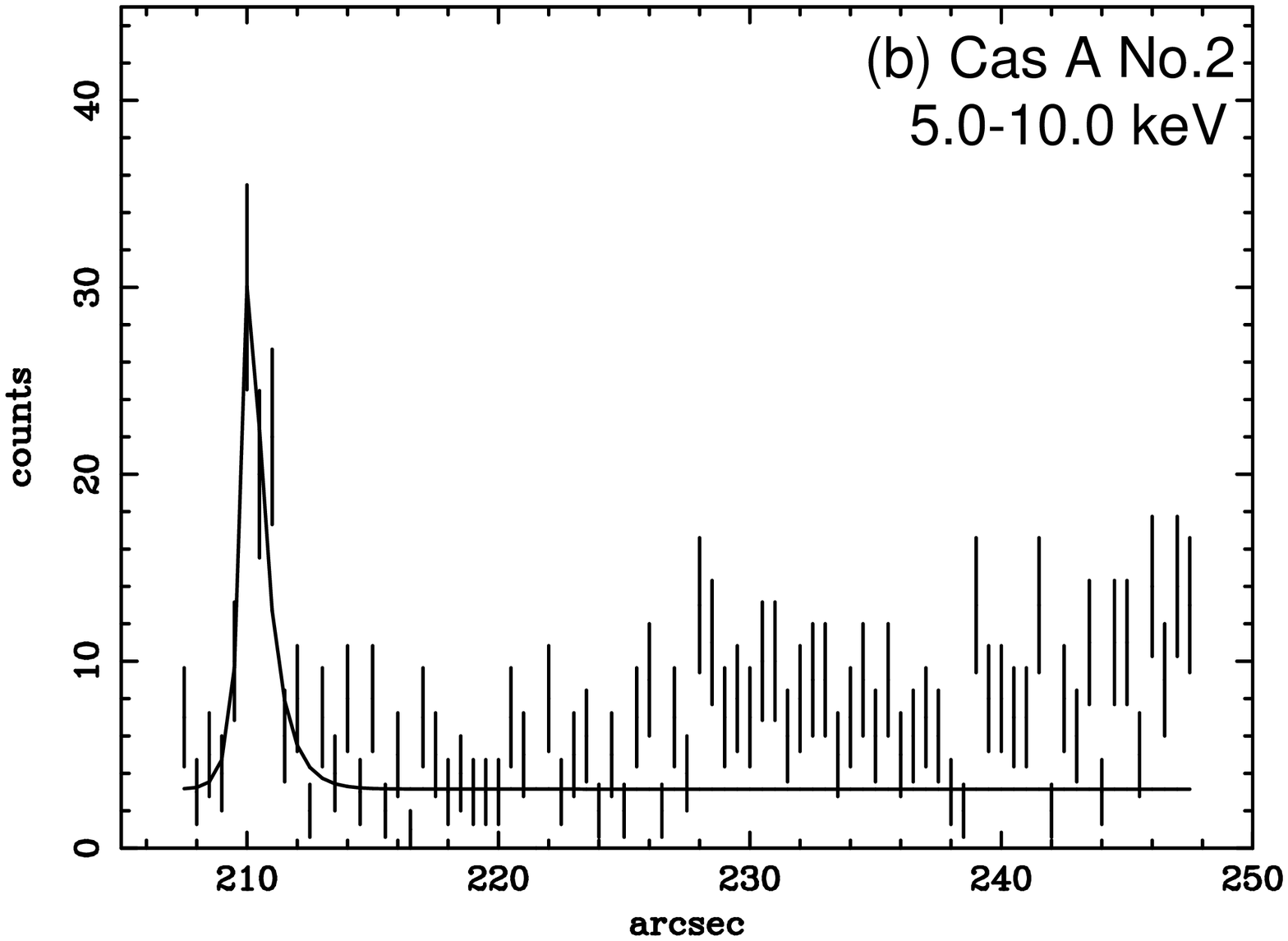}
\plotone{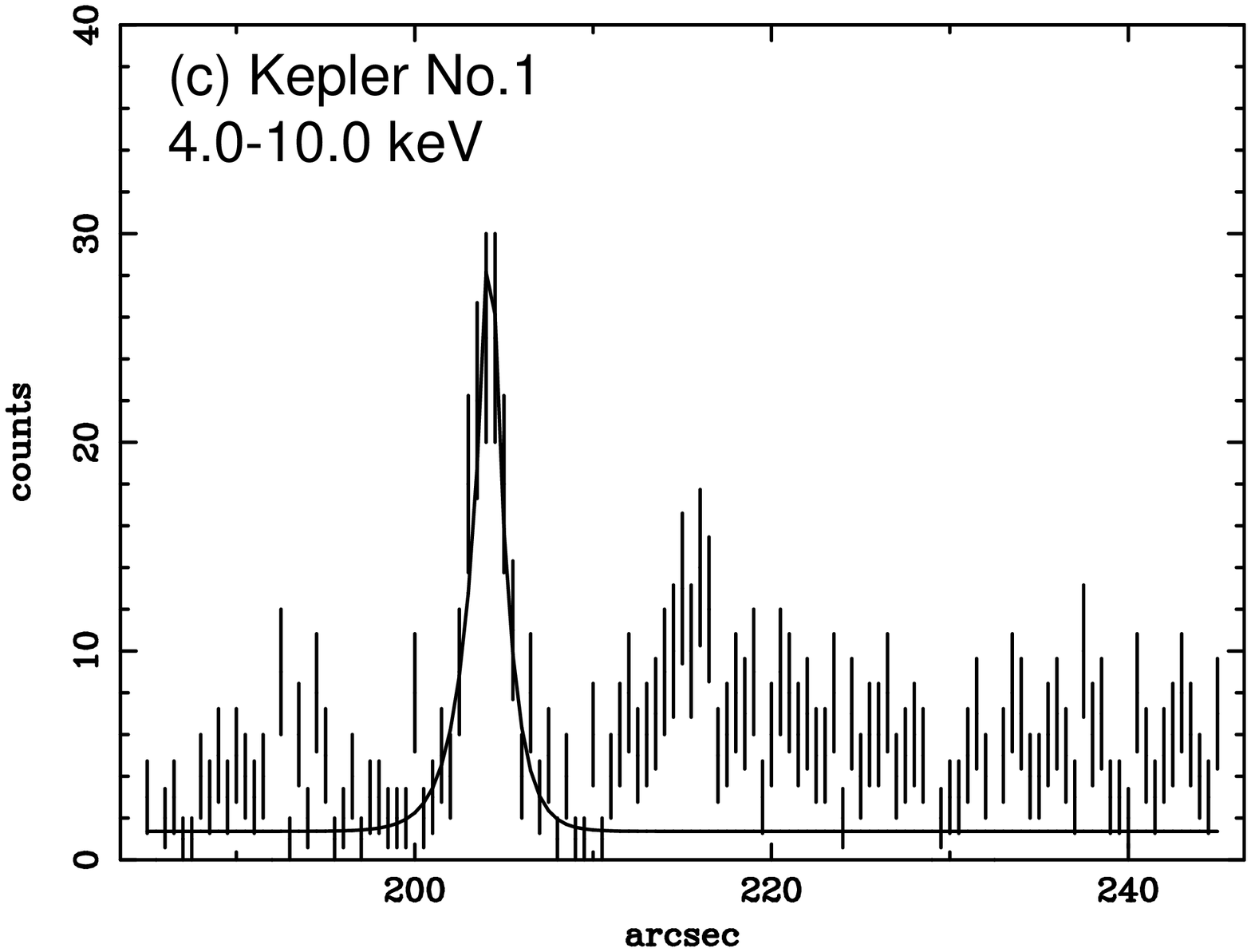}
\plotone{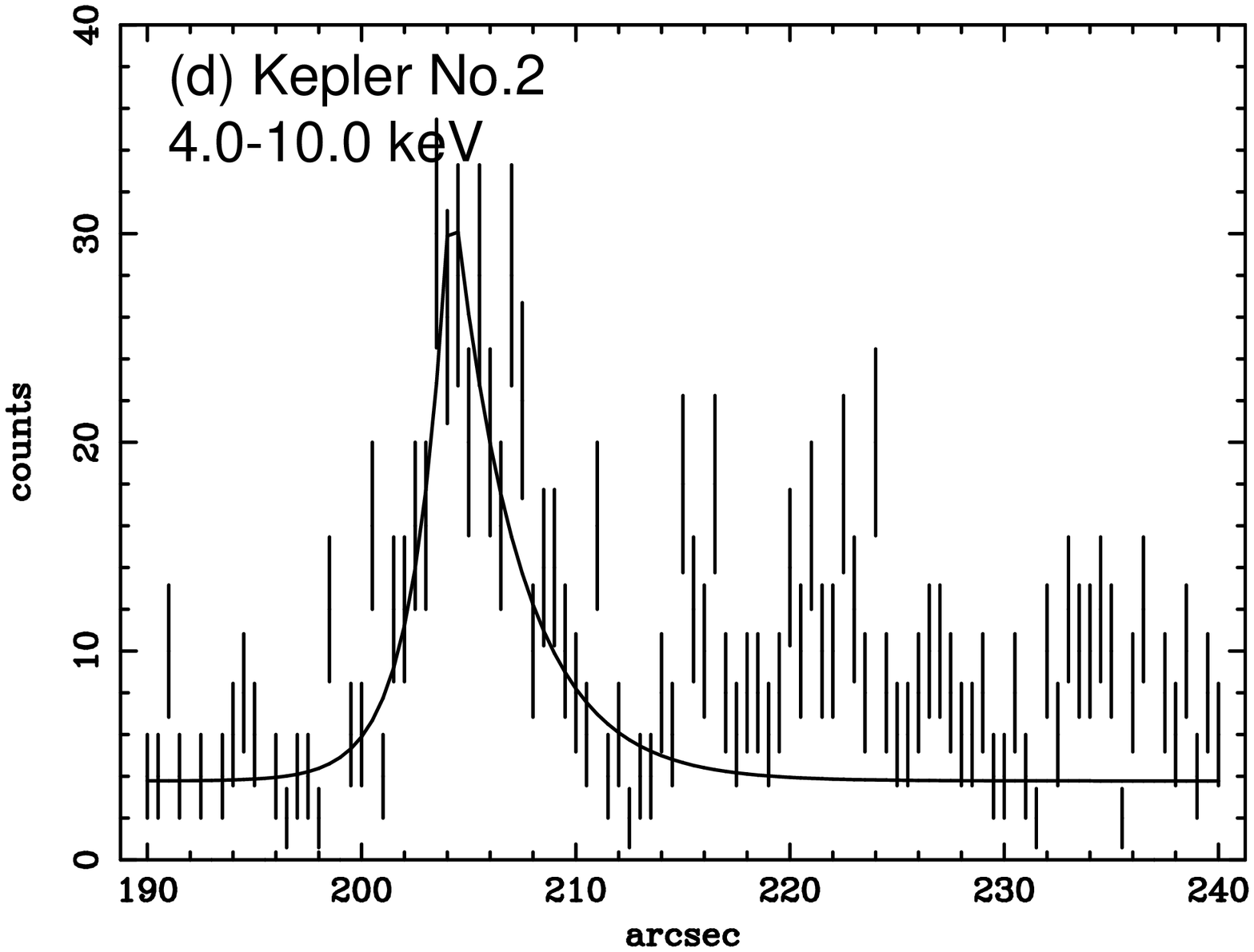}
\plotone{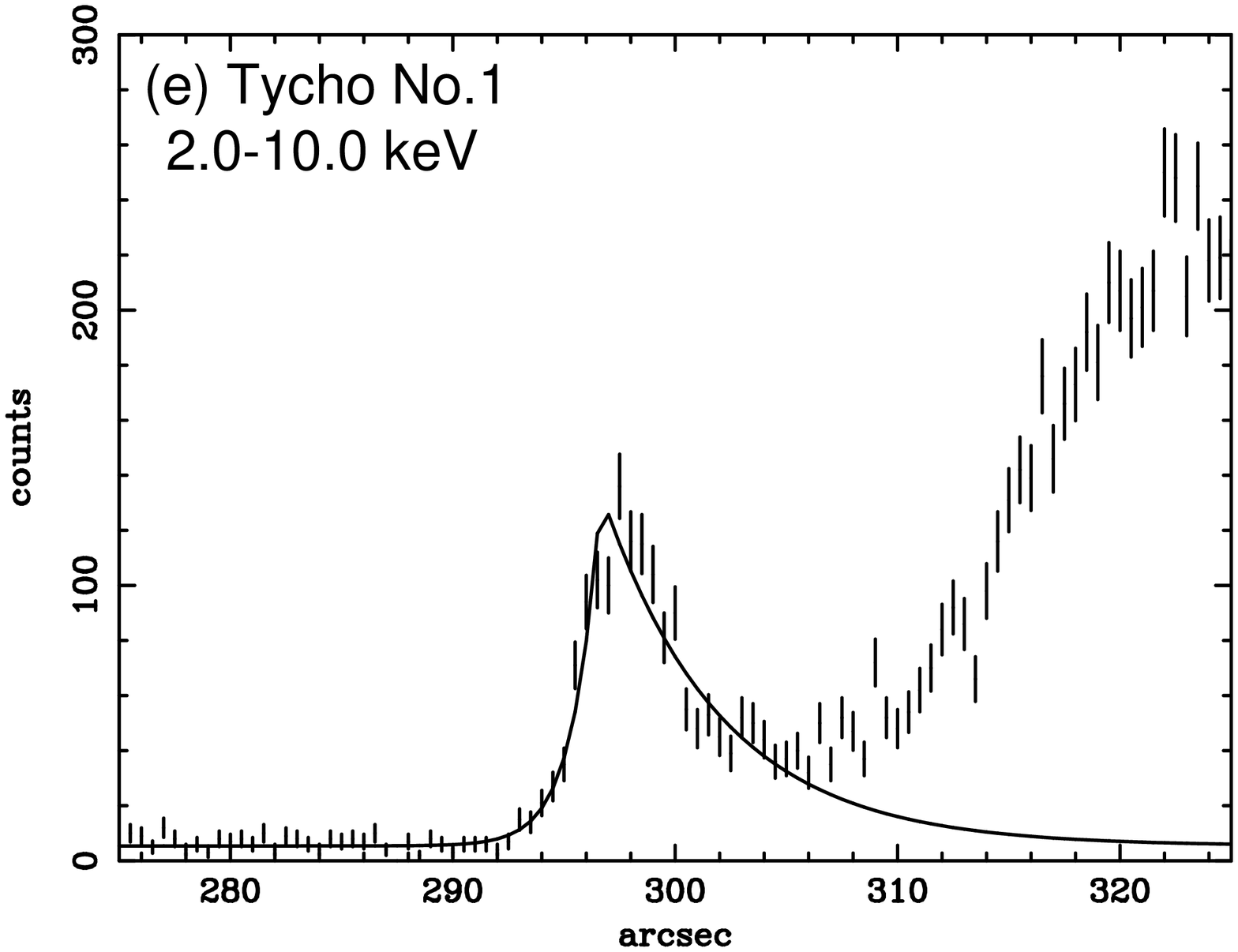}
\plotone{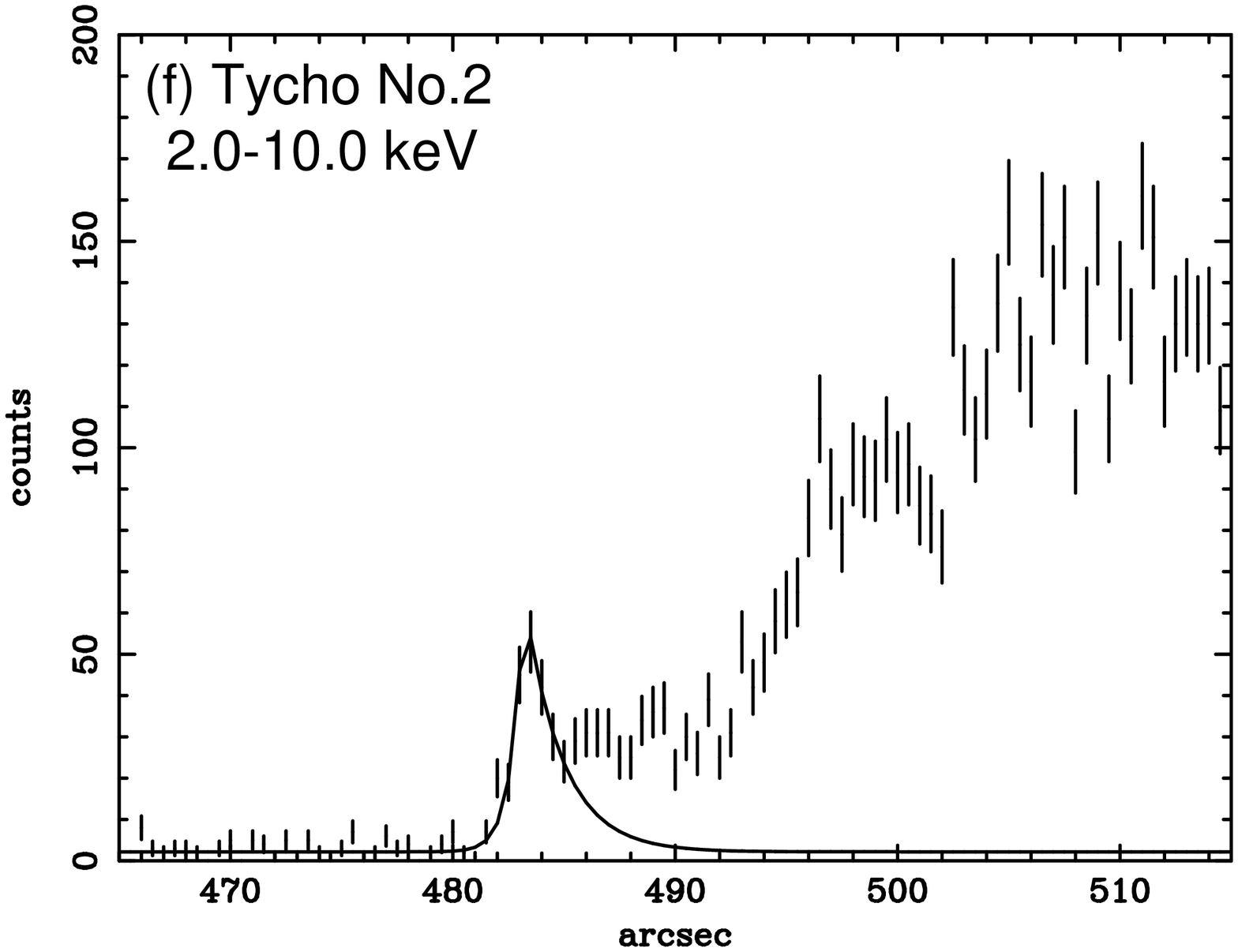}
\plotone{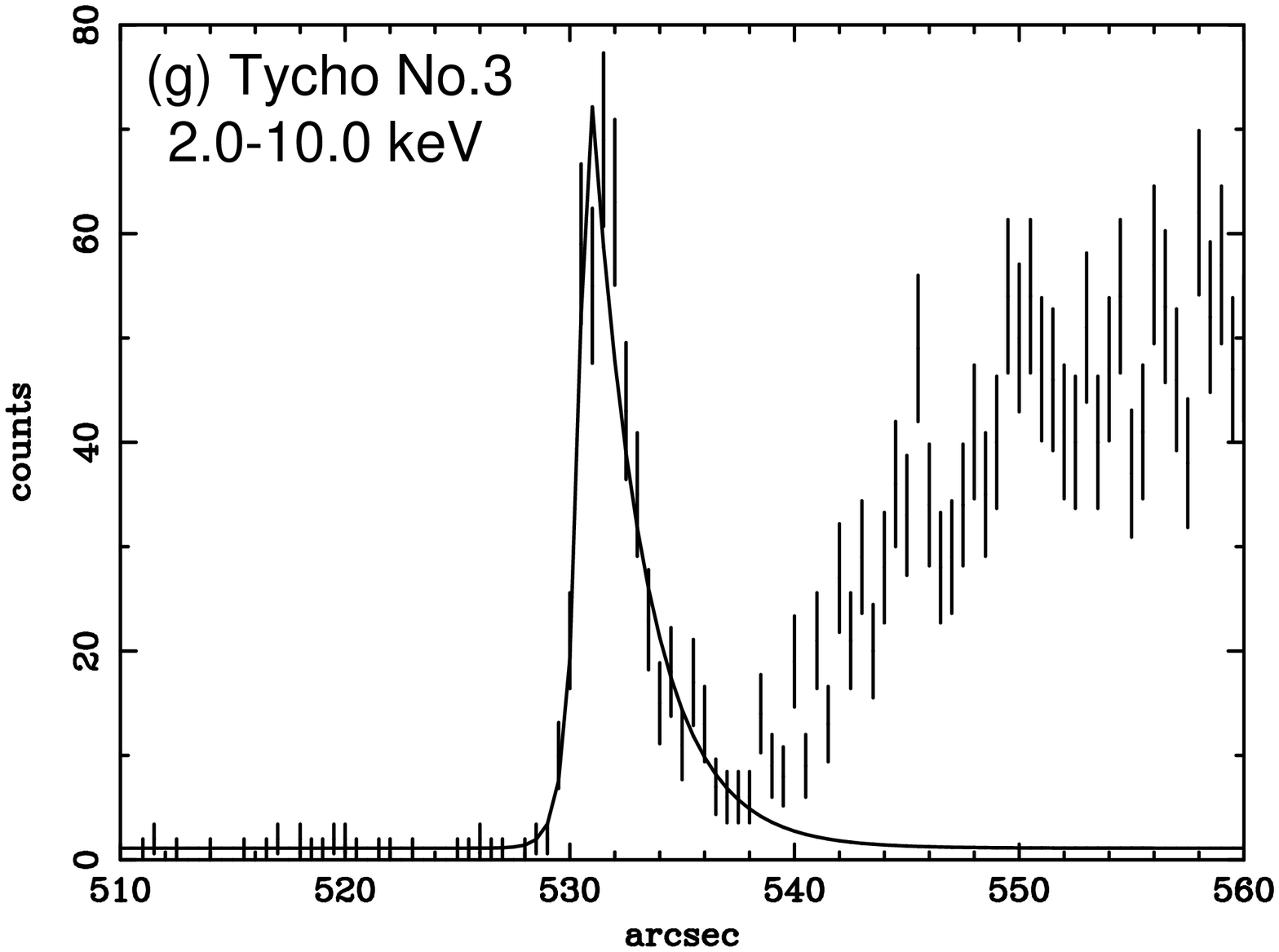}
\plotone{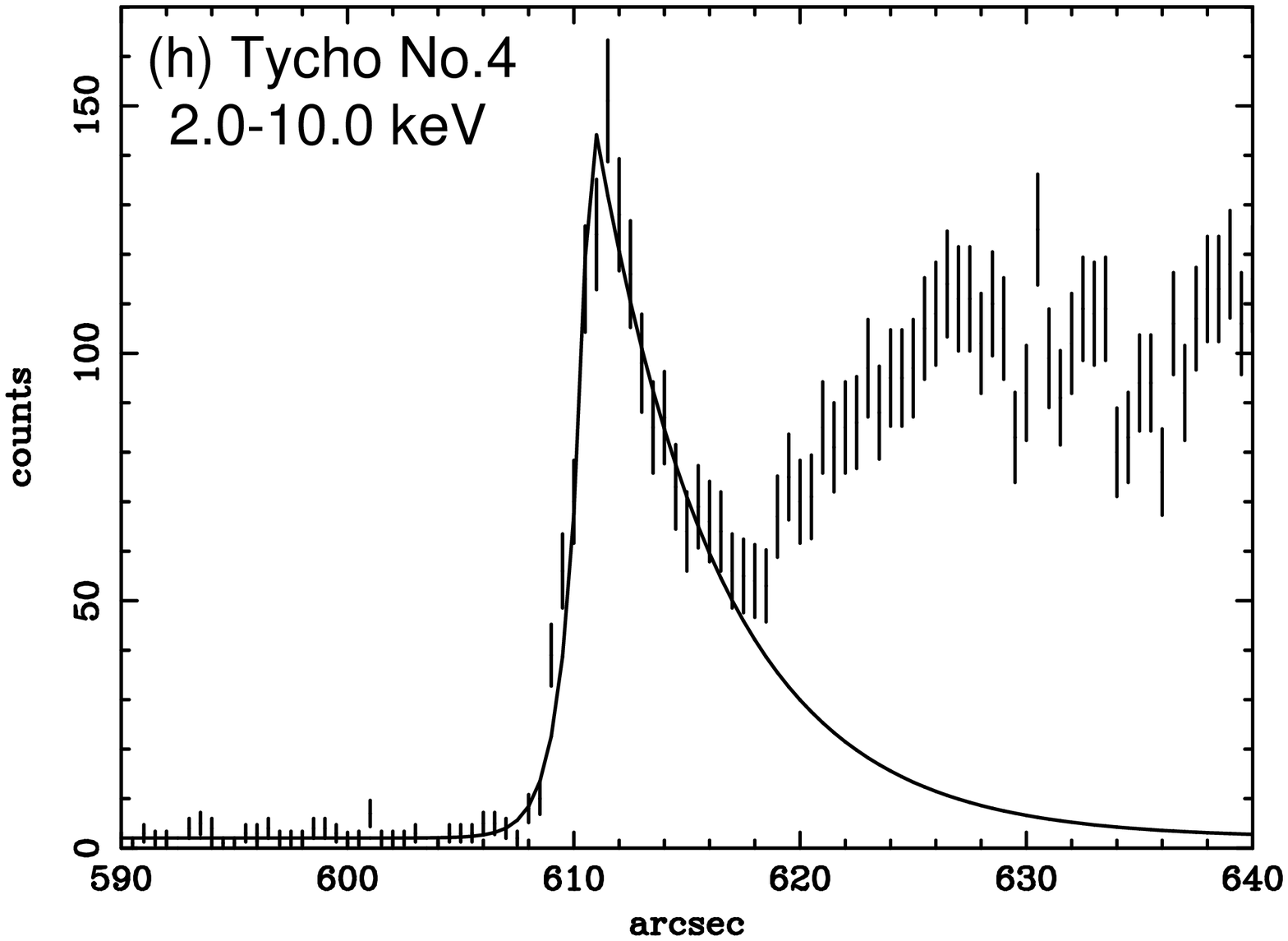}
\plotone{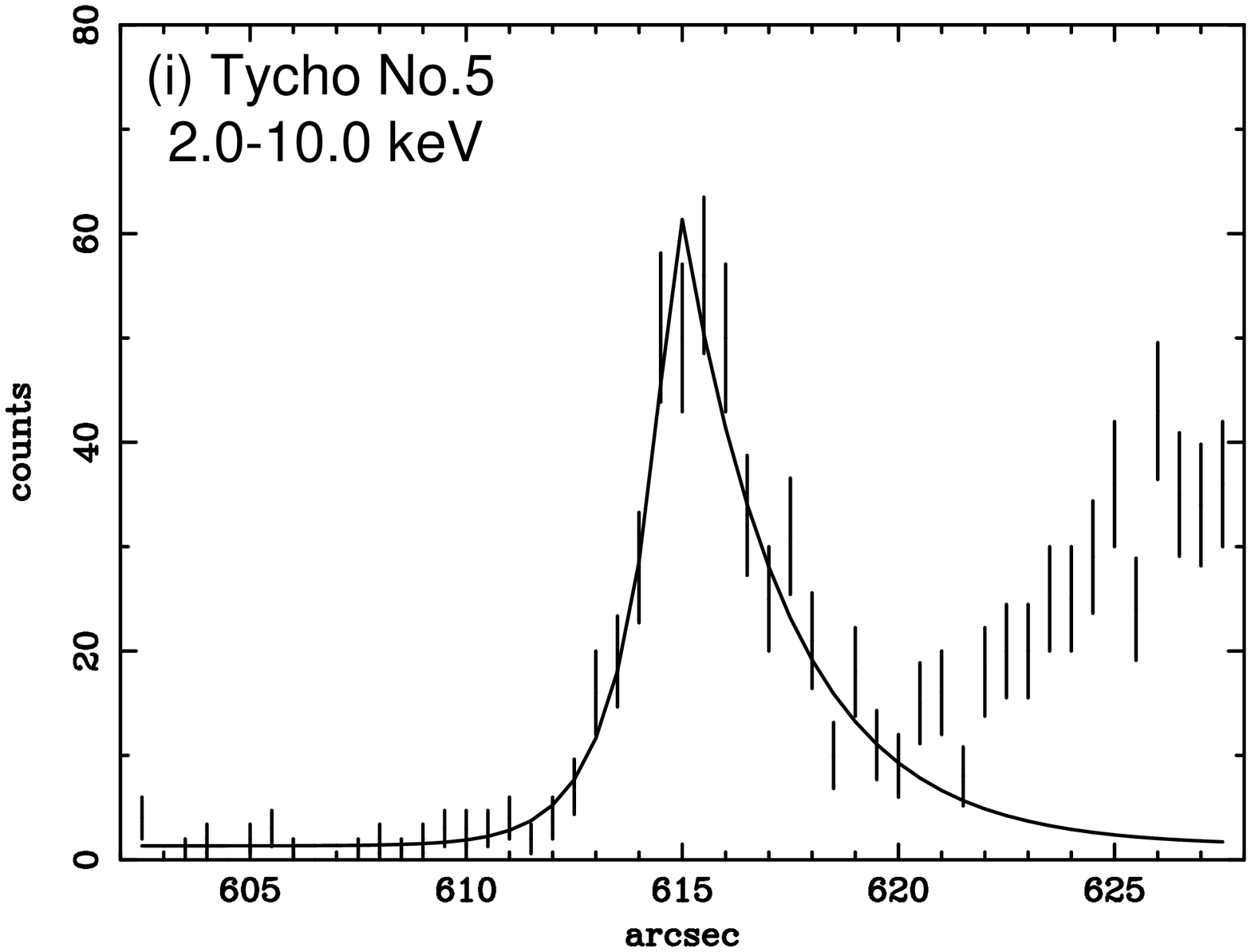}
\plotone{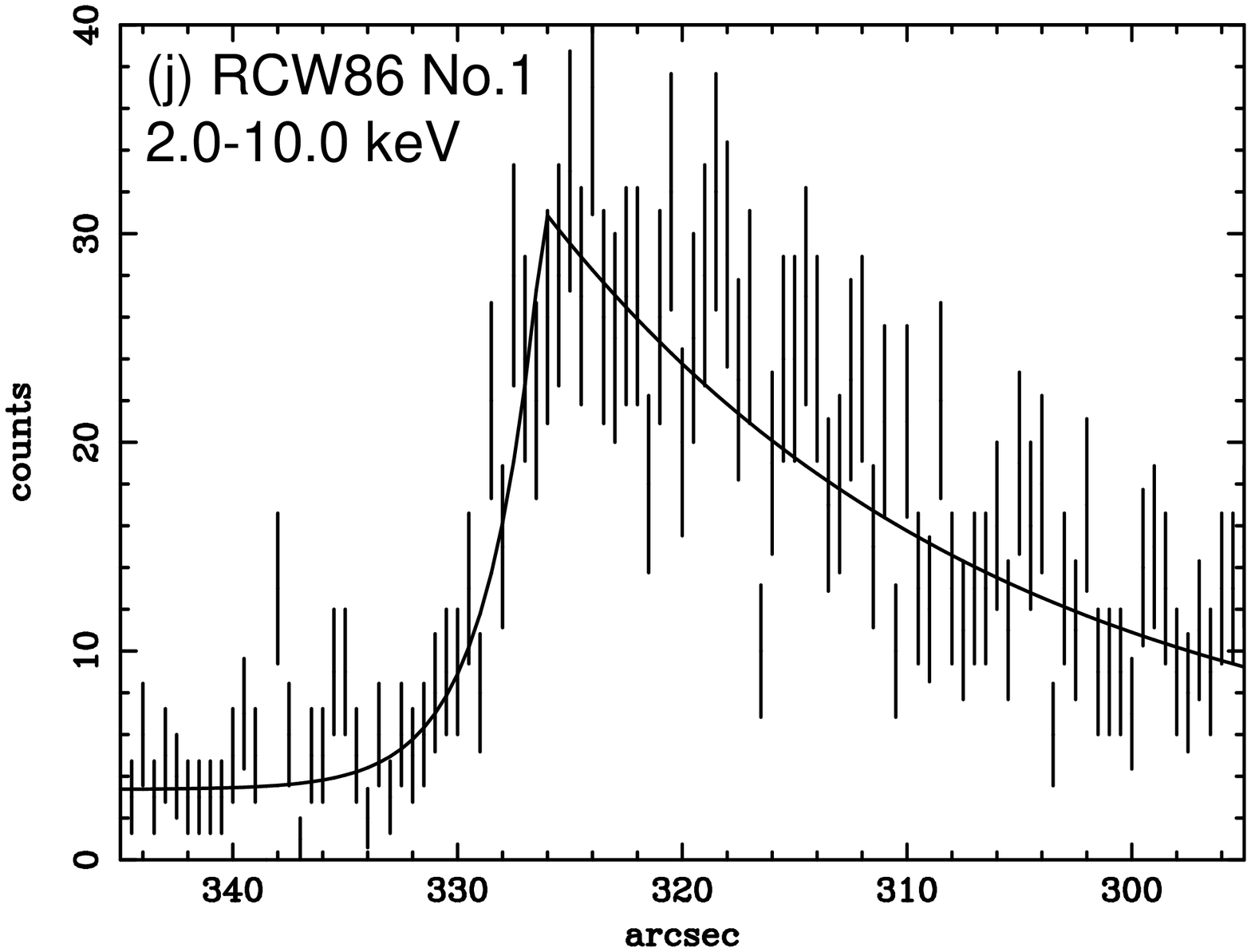}
\plotone{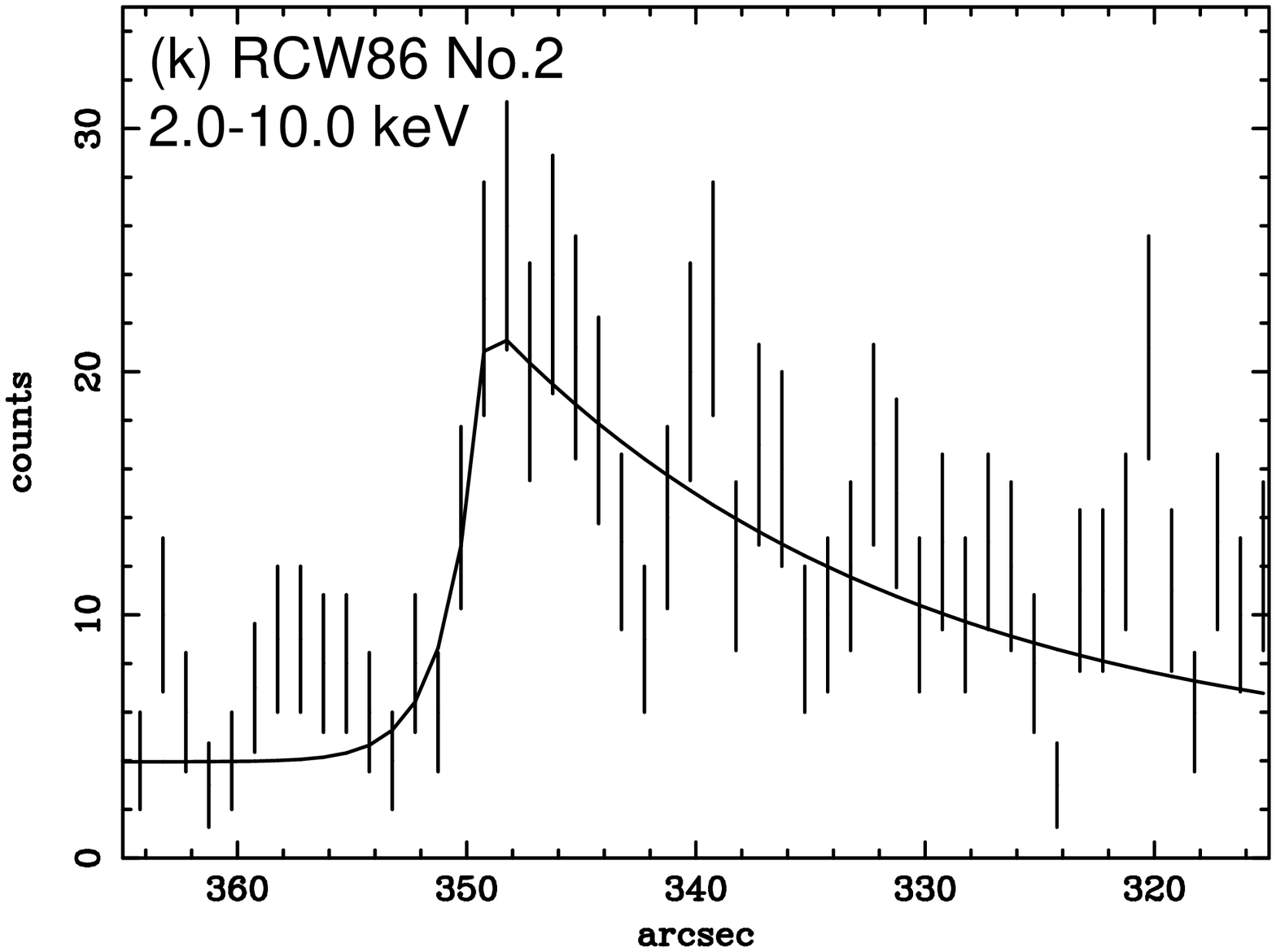}
\caption{Profiles of the filaments of SNRs in the hard X-ray band
(Cas~A: 5.0--10.0~keV, Kepler: 4.0--10.0~keV, 
Tycho: 2.0--10.0~keV, RCW~86: 2.0--10.0~keV).
The best-fit models are shown with solid lines.
The shock runs from right to left in the all panels.
Note that fittings were carried out in a part around the shock front
in order to avoid any contamination from bright thermal emission.}
\label{fig:profiles}
\end{figure}

Figures~\ref{fig:profiles}(a) and (b) show
the profiles of two filaments.
In order to estimate the scale width of these filaments,
we used a simple, empirical fitting model
for the apparent profiles,
which is the same
as used in the previous analysis in the SN~1006 case by \citet{bamba2003b}:
\begin{equation}
f(x) = \left\{
        \begin{array}{rlr}
        A\exp \left(-|\frac{x_0-x}{w_u}|\right) & {\rm on\ the\ upstream\ side} \\
        A\exp \left(-|\frac{x_0-x}{w_d}|\right) & {\rm on\ the\ downstream\ side},
        \end{array}
\right.\label{model}
\end{equation}
where $A$ and $x_0$ are the flux and position at the emission peak,
respectively.
The quantities $w_u$ and $w_d$ represent the e-folding widths on
the upstream and the downstream sides of the apparent emission peak, 
respectively
(hereinafter, "$u$" and "$d$" represent
upstream and downstream sides, respectively).
We treat the best-fit values smaller than 0.8~arcsec as the upper limit,
because they highly suffer the influence of point spread function (PSF),
which is about 0.5 arcsec
within $\sim$4~arcmin radius from the aim point.
The fittings were made with profiles in the 5.0--10.0~keV,
and were accepted statistically
with the best-fit models and parameters
shown in Figures~\ref{fig:profiles}(a) and (b), and Table~\ref{tab:fila_image}.

We then made the spectra of the filaments within the
scale widths
($x_0-w_u \leq x \leq x_0+w_d$)
in Figures~\ref{fig:profiles}(a) and (b).
The background spectra were made from 
the just off-filament downstream sides with the same width of the filaments.
We compared the Si line photon counts 
with those in the continuum (5.0--10.0~keV) band
in these regions,
and found that photons in this band include
thermal photons in a ratio of only less than 25\%,
under the assumption that
all photons in the background regions are thermal,
and both regions have a thermal component
with the same physical parameters, except for their flux.
Therefore, we treated all the 5.0--10.0~keV photons
in our analysis of Cas~A,
as a part of the hard X-rays
reported in previous observations \citep{vink2003}.

\begin{figure}[hbtp]
\epsscale{0.45}
\plotone{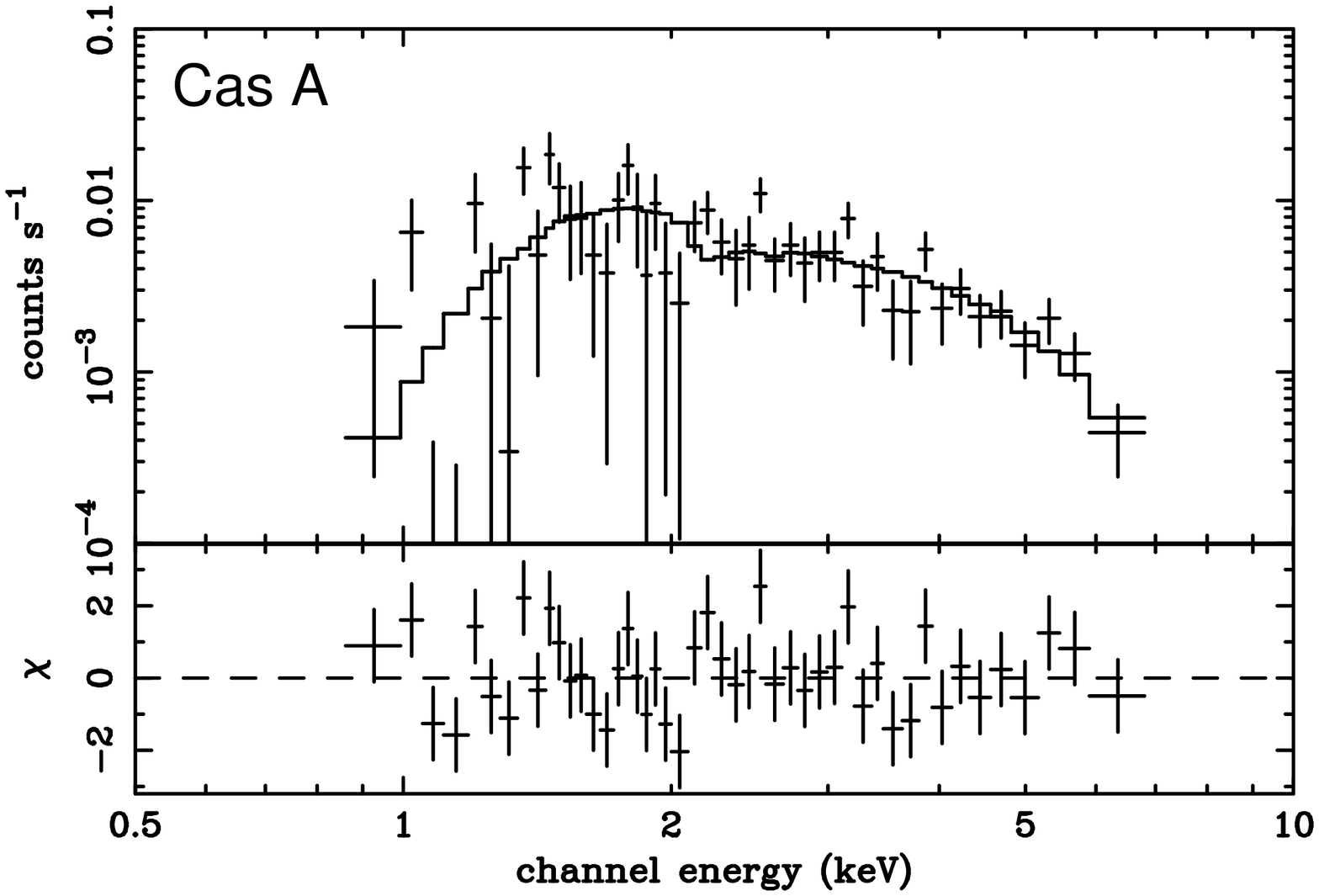}
\plotone{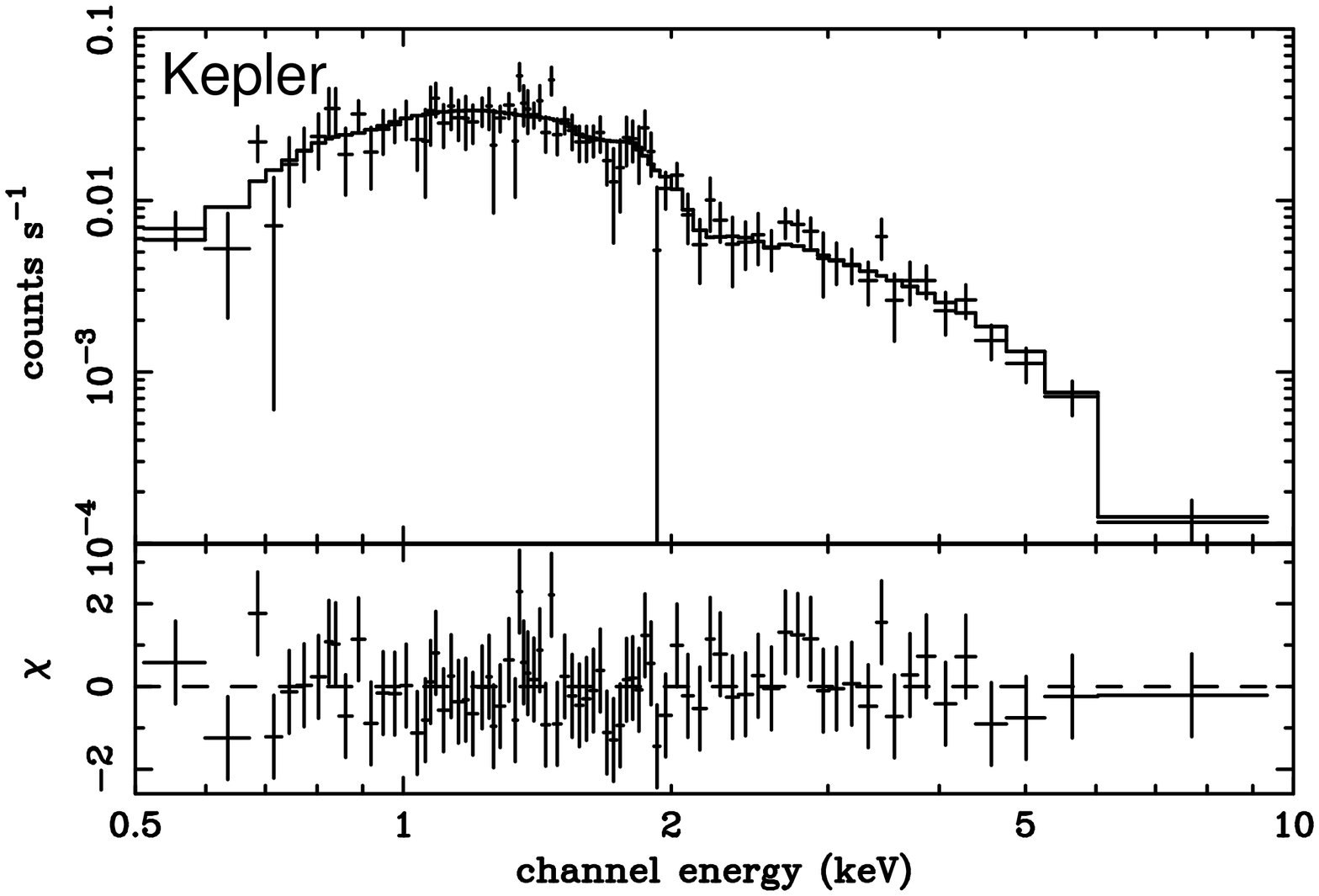}
\plotone{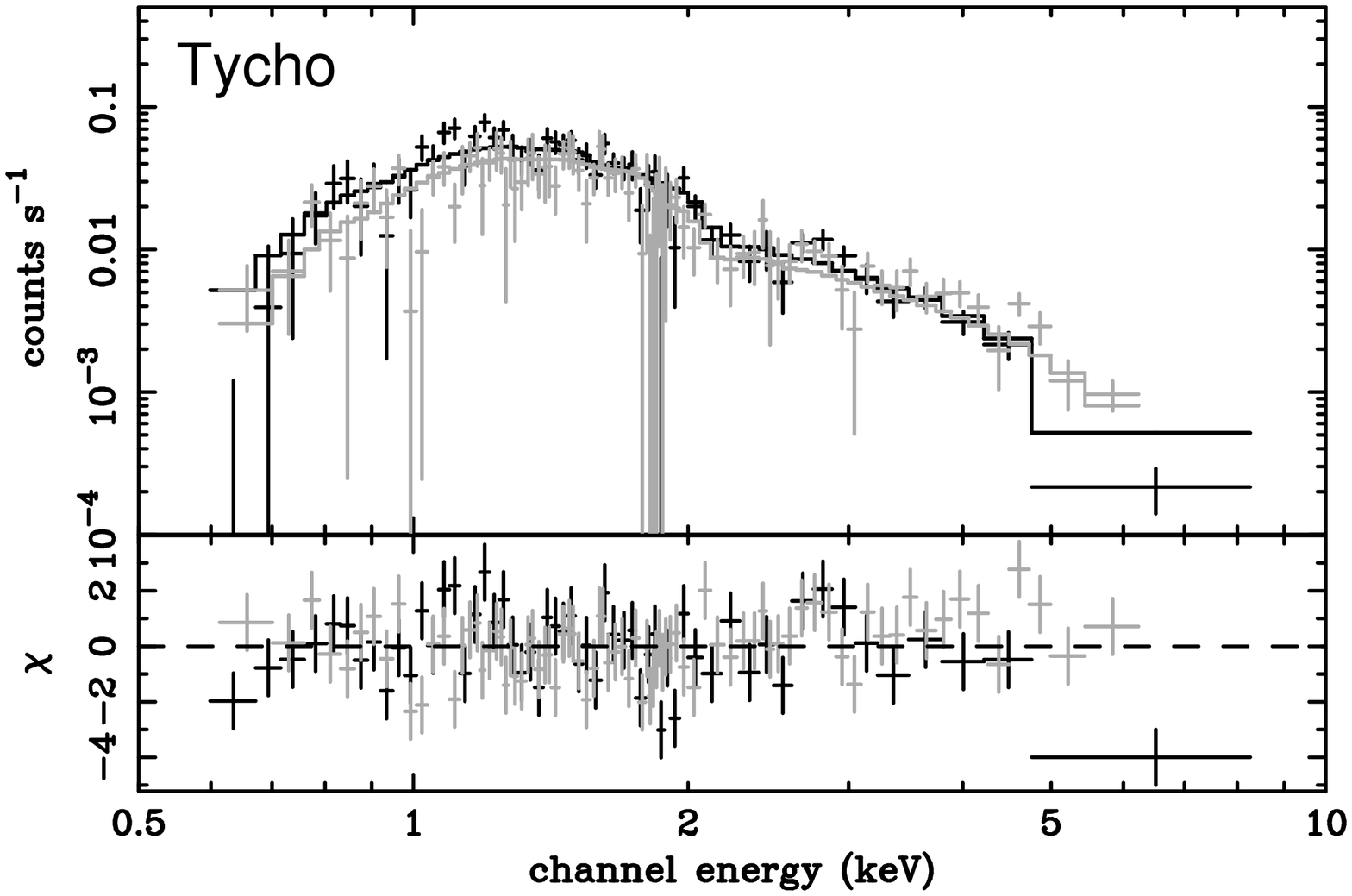}
\plotone{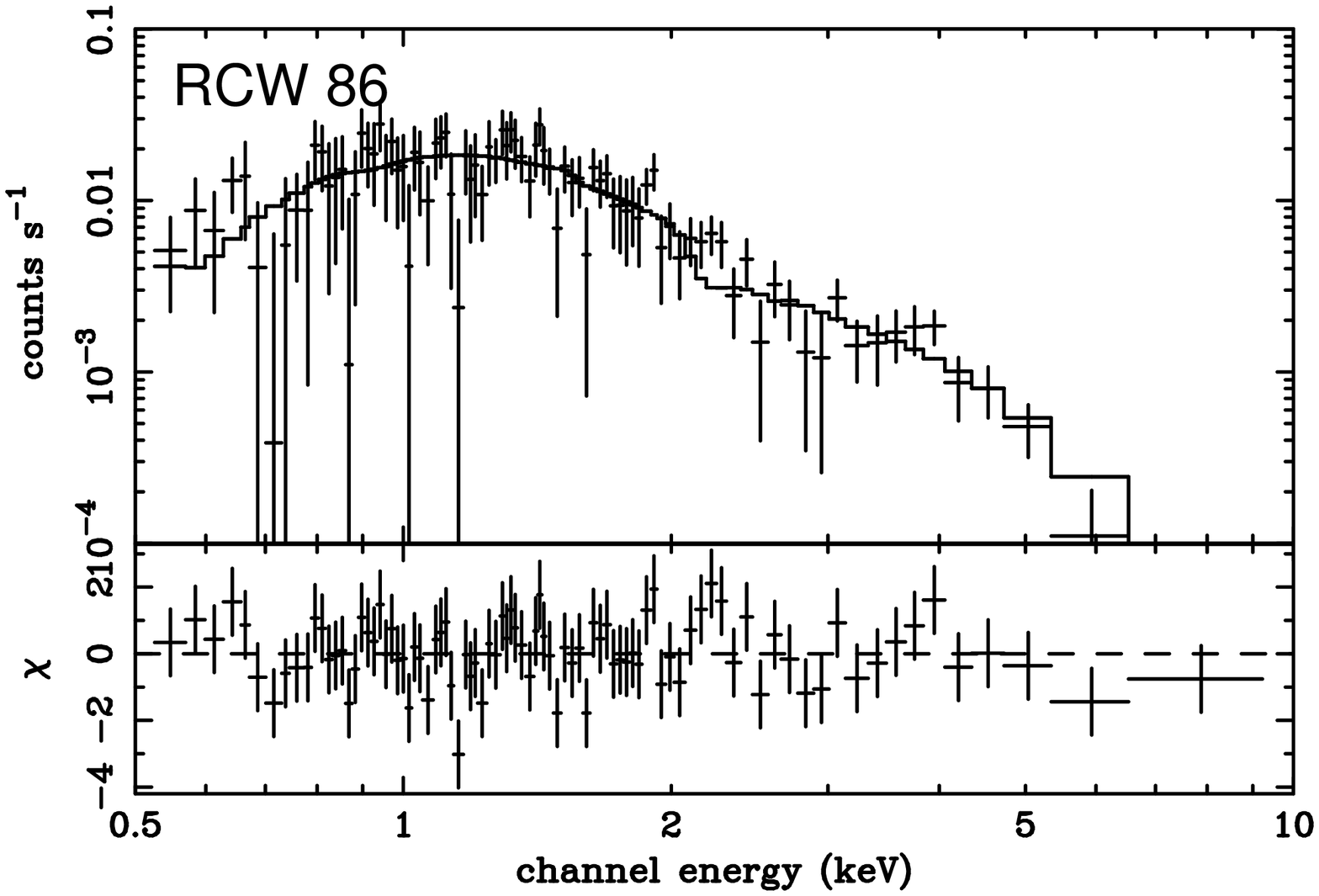}
\caption{Spectra of combined-filaments for each SNR (crosses).
The best-fit power-law models are shown in the solid lines.
The black and gray crosses in the panel for Tycho
are the spectra accumulated from filaments on the different CCD chips.
The lower panel for each figure represents the residuals
from the best-fit model.}
\label{fig:spectra}
\end{figure}

The background-subtracted spectrum accumulated from the two filaments
is shown in the upper-left panel of Figure~\ref{fig:spectra},
which is featureless with no line-like structure,
and extends to the hard X-ray side,
which was fitted with both an absorbed thin thermal plasma model
in non-equilibrium ionization (NEI)
calculated by \citet{borkowski2001a}
("{\tt NEI}" model),
and an absorbed power-law model.
The absorption column was subsequently calculated using the cross sections by
\citet{morrison1983} with the solar abundances \citep{anders1989}.
Both models were well fitted statistically with best-fit curves and parameters,
shown in Figure~\ref{fig:spectra} and Table~\ref{tab:fila_spec}.
We also applied the {\tt SRCUT} model,
which represents the synchrotron emission from electrons of power-law
distribution with exponential roll-off in a homogeneous magnetic field
\citep{reynolds1999}.
The spectral index at 1~GHz was fixed to be $0.78$
according to a report by \citet{green2004}.
This model also well reproduces the spectra
with best-fit values shown in Table~\ref{tab:fila_spec}.
The spectra are much harder than those of the filament analyzed by
\citet{vink2003}.

\subsection{Kepler's Remnant (G4.5+6.8)}

Kepler's remnant is the second youngest SNR in our Galaxy
recorded by human beings,
which appeared in 1604 \citep{kepler1606}.
\citet{delaney2002} observed this SNR with VLA,
and found the spatial variation of spectral index from $0.85$ to $0.6$
with a mean value of $0.7$.
The distance to the remnant is $4.8\pm1.4$~kpc
using the H~I observations with VLA \citep{reynoso1999}.
\citet*{petre1999} found hard X-rays (above 10~keV) from Kepler
with {\it RXTE},
and {\it XMM-Newton} pointed out that
at least the south-eastern edge emits nonthermal X-rays
\citep{cassamchenai2004}.

The upper-right panel of Figure~\ref{fig:images} shows
a 4.0--10.0~keV band {\it Chandra} image of the eastern rim of Kepler.
We can see very sharp structures on the outer edge of the remnant,
especially bright on the eastern side,
which is similar to the filaments in SN~1006.
In order to carry out spatial and spectral analysis of these structures,
we selected two filaments, as shown in the panel,
which are $<$160~arcsec distant from the aim-point.
Figures~\ref{fig:profiles} (c) and (d) show
the profiles of the filaments
in the 4.0--10.0~keV band.
They have sharp edges both on the upstream and downstream sides of
the emission peak,
and eq.(\ref{model}) well represents their profiles 
with best-fit curves and parameters 
summarized in the panels and Table~\ref{tab:fila_image}.

We obtained the spectra of these filaments
in the same way of the Cas~A case.
Background regions were selected just
from the downstream sides of the filaments,
which are free from any other structures.
The intensities of iron and silicon lines and continuum photons 
in the 4.0--10.0~keV band were compared,
and it was found that
more than 90\% of the photons in the 4.0--10.0~keV band
are the hard X-ray component 
reported in previous observations \citep{petre1999,cassamchenai2004},
with the same assumptions as in the Cas~A case.
Thus the derived scale widths represent those of the hard X-ray emission.
The background-subtracted spectrum accumulated from these filaments
is shown in the upper-right panel of Figure~\ref{fig:spectra},
which shows a hard and line-less feature.
We fitted each spectrum 
with {\tt NEI} and power-law models
with an absorption.
Both models are well fitted to the data
with the best-fit models and parameters
shown in the panel and Table~\ref{tab:fila_spec}.
The {\tt SRCUT} model was also applied
with the spectral index at 1~GHz fixed to be $0.70$,
derived by \citet{delaney2002} as the average value on the eastern side.
This model also well represents the data
with the best-fit parameters listed in Table~\ref{tab:fila_spec}.

\subsection{Tycho's Remnant (G120.1+1.4)}

Tycho is a remnant of a supernova 
that exploded in 1572 \citep{tycho1573}.
The distance to the remnant is estimated to be 1.5--3.1~kpc,
from the proper motion of the optical filaments \citep{kamper1978}
and the shock velocity estimated by \citet{ghavamian2001}.
We adopt the most popular value, 2.3~kpc.
\citet{katzstone2000} investigated
the variations of the radio spectral indices,
and found that the emission in the outer rim
shows a trend that brighter clumps have a flatter spectral index
than the average index of $s = 0.52\pm 0.02$,
maybe due to either SNR blast waves and ambient medium interactions or
internal inhomogeneities of the magnetic field within the remnant.

Hard X-ray emission has been unmistakably detected from Tycho,
up to 30~keV,
with {\it HEAO 1} \citep{pravdo1979},
{\it Ginga} \citep{fink1994}, 
and {\it RXTE} \citep{petre1999}.
{\it Chandra} and {\it XMM-Newton} also observed Tycho's remnant,
and the good spatial resolutions of instruments onboard these satellites
show the spatial variety of X-ray emission
\citep{hwang2002,decourchelle2001}.
\citet{hwang2002} claimed that 
there is a thin structure encircling the SNR,
the spectrum of which has lines with low equivalent widths.

In order to be clear as to
whether the thin structure is a hard X-ray emitter or not,
spatial and spectral analysis were conducted in the same way 
as in the cases for Cas~A and Kepler.
The lower-left panel of Figure~\ref{fig:images} shows
the north-western part of the remnant in the 2.0--10.0~keV band.
The thin filament-like structure suggested by \citet{hwang2002}
can be seen.
Figure~\ref{fig:profiles} (e)--(i) shows
profiles of the filaments selected in the panel,
which are located within 120~arcsec from the aim-point.
Very sharp rises and decays can be seen in all of the filaments.
The fittings with eq.(\ref{model}) were performed only
around the outermost peak of hard X-ray emission,
because thermal photons dominate in the downstream regions
(the right side of each panel),
and were accepted statistically
with the best-fit models and parameters shown
in these panels and Table~\ref{tab:fila_image}.

Their spectra were obtained in the same way as in the Cas~A case.
Background spectra were accumulated from just the downstream side
of the filaments.
The lower-left panel of Figure~\ref{fig:spectra} shows
the background-subtracted spectra,
which were combined with those of filaments in each CCD chip.
Since both are hard and have no line-like structure,
we fitted them with {\tt NEI}, power-law, and {\tt SRCUT} models
with an absorption.
We fixed the spectral index at 1~GHz to be $0.52$,
following \citet{katzstone2000}.
All three models were accepted statistically
with similar probability,
with the best-fit values listed in Table~\ref{tab:fila_spec}.
We compared the Si and S lines and the 2.0--10.0~keV continuum band intensity
in the source spectrum with those in background region,
and it was found that
only less than 6\% of the thermal photons contribute in the 2.0--10.0~keV band.
Therefore,
it is likely that
the all photons in the 2.0--10.0~keV band can be treated as pure nonthermal.

\subsection{RCW~86 (G315.4$-$2.3)}

RCW~86 was identified as 
being the oldest historical SNR by \citet{clark1977}.
They suggested that
a progenitor of the remnant was reported in AD~185
in Later Han dynasty records \citep{fanyou432},
although \citet{chin1994} claimed some doubt
about the record.
The spectral index at 1~GHz and the distance was measured
to be $0.6$ by \citet{caswell1975} and 2.8~kpc \citep{rosado1996},
respectively.

Strong hard X-rays have been detected from the remnant
by {\it Ginga} \citep{kaastra1992} and {\it RXTE} \citep{petre1999}.
Using {\it ASCA} data,
\citet{bamba2000} and \citet{borkowski2001b} independently found that
the hard X-rays show a nonthermal feature,
and concentrate on the south-western (SW) shell of the SNR.
The {\it Chandra} observation with excellent spatial resolution
reveals that
the regions emitting nonthermal X-rays are very clumpy
and different from those emitting thermal X-rays \citep{rho2002}.

The lower-right panel of Figure~\ref{fig:images} represents 
a close-up view of the SW shell in the 2.0--10.0~keV band.
Clumpy structures can be seen,
which have been already indicated as nonthermal filaments by \citet{rho2002}.
We selected two filaments for our analysis,
which are bright in the hard band image
and are located $<$120~arcsec from the aim-point.
Their profiles from the region indicated in the panel were made,
as shown in Figures~\ref{fig:profiles}(j) and (k).
They show a sharp rise and a rather slow decay in the upstream and
the downstream sides of the emission peak,
respectively.
The fitting with eq.(\ref{model}) was carried out in the 2.0--10.0~keV band,
and was accepted statistically with best-fit models and the values
shown in the panels and Table~\ref{tab:fila_image}.

The spectra were made in the same way as in the former cases.
The background regions were selected from the downstream side of the filaments
without other structures.
The emission is free from thermal photons,
according to previous studies,
especially in such a hard band \citep{rho2002}.
The lower-right panel of Figure~\ref{fig:spectra} shows
the background-subtracted spectrum,
which was accumulated from both filaments.
The spectral fittings were made with the three models
({\tt NEI}, power-law, and {\tt SRCUT}) with an absorption.
We adopted the spectral index at 1~GHz 
with the value  $0.6$ derived by \citet{caswell1975}.
All of the models well reproduced the data
with the best-fit profiles and values given
in the panel and Table~\ref{tab:fila_spec}.

\section{Discussion}
\label{sec:discussion}

\subsection{Emission Origin of Filaments}
\label{sec:origin}

Table~\ref{tab:fila_spec} summarizes the spectral fittings
for the filaments in the four SNRs
to thermal ({\tt NEI}) and nonthermal (power-law and {\tt SRCUT}) models.
Because all models give a similar reduced $\chi^2$,
we cannot distinguish which model is the best statistically
to reproduce the spectra of the filaments.
The thermal model fittings for all cases, however, require
an unusually high temperature and a low abundance
for young (ejecta dominated) SNRs.
On the other hand,
the best-fit photon indices in the power-law model are $\sim$2--3,
which are similar to hard X-rays in other SNRs
\citep[e.g.,][]{koyama1995}.
It reminds us of the synchrotron emission from electrons 
with a power-law distribution of index of $\Gamma=1.5$ or softer,
suggesting that these filaments are acceleration sites of electrons
via the DSA mechanism.
The fitting of the {\tt SRCUT} model requires
a roll-off frequency of about 1~keV,
as shown in Table~\ref{tab:fila_spec}.
In order to check the influence of uncertainty of the radio index $s$,
we fitted the spectra with the frozen $s$ of 0.5 and 0.9.
For the $s=0.5$ cases, 
which is the theoretical minimum indicated by the DSA,
$\nu_{rolloff}$ remains unchanged within a factor of two.
On the other hand, the fittings with $s=0.9$,
which implies that the de-acceleration is very efficient,
require the change of $\nu_{rolloff}$ in the order of 1.
However, the best-fit normalization in the radio band becomes 
larger than 10\% of the total flux from the whole SNR,
which is unrealistic for such small regions.
Moreover, in the Kepler case, the radio map shows that 
the value of $s$ of the rim is smaller than
that of the inner region \citep{delaney2002},
although it changes from 0.85 to 0.6.
This result suggests that the outer rims of SNRs
are acceleration sites and de-acceleration may not occur yet,
in other words, $s$ is around 0.6.
For the Tycho case, 
\citet{katzstone2000} indicated that
the spectral index of filaments in outer rim is around 0.5,
indicating that $\nu_{rolloff}$ derived in our analysis for the SNR
is roughly correct.
These results indicates that
the X-ray photons shown in Figure~\ref{fig:images} are emitted by
electrons with the highest energy in each SNR.

It can be argued that the thermal spectra in source and background
regions may be the same.
The line emission from the NEI plasma changes
as the ionization time scale ($n_et$) varies.
We take the background regions just behind the source regions,
and their widths are much small (1--5\% of SNR radii).
The electron density $n_e$ may not change so drastically
in such thin downstream regions.
Since we take the background regions just behind the source regions,
and their widths are much small (1--5\% of SNR radii),
the electron density $n_e$ does not so drastically change
in the downstream regions.
Thus, the difference of $n_et$ between the background and source regions
may not be large
enough to clarify the difference between their thermal spectra. 

All of the filaments have much smaller scale widths
than that derived from the Sedov self-similar solution ($=R/12$),
on both the upstream and downstream sides,
which is similar to the previous results in the SN~1006 case
\citep{bamba2003b}.
We should consider the influence of PSF,
especially for the filaments in Cas~A and Tycho,
where $w_u$ are as small as the PSF (see Table~\ref{tab:fila_image}).
The PSF enlarges the apparent $w_u$ and $w_d$.
Thus, our result may be the upper limit of the physical scale width.
Together with the fact that
our sample covers almost historical SNRs with an X-ray shell,
it may be suggested that
all relative young ($\la$2000~years) SNRs accelerate electrons
in very thin regions around shock fronts.

\subsection{Comments on Individual SNRs}
\label{sec:individual2}

\subsubsection{Cas~A}

We analyzed filaments thinner than that by \citet{vink2003},
and the former has harder spectra than the latter.
This fact may imply that
the magnetic field around our filaments is stronger than that around
the filament by \citet{vink2003}.
Based on analysis of the filament-like structure in Cas~A by
\citet{vink2003},
\citet{berezhko2004}
estimated the magnetic field around the structure
to be $\sim 500~\mu$G,
generated by the cosmic rays themselves \citep{lucek2000}.
Our thinner filaments may have
a larger magnetic field according to their discussion.

\subsubsection{Kepler's Remnant}

As for Kepler, since there is no report about nonthermal filaments,
this is the first analysis of them.
\citet{cassamchenai2004} reported that
south-eastern edge of the SNR emits nonthermal X-rays
with {\it XMM-Newton}.
Their spectral parameters are consistent with ours.
Therefore, the region by \citet{cassamchenai2004}
may be a part of our filaments.

\subsubsection{Tycho's Remnant}

Our spectral results are roughly consistent with
the results by \citet{hwang2002},
although the regions for spectral analysis are slightly different,
except for the S abundance in the {\tt NEI} model.
The difference may come
mainly from selection of the different background regions;
we chose them not from out of the SNR, but from just at the downstream side
of the filaments.
We could thus extract pure nonthermal emission.

\subsubsection{RCW~86}

We re-confirmed the result of spectral analysis by \citet{rho2002}.
A similar discussion about the magnetic field in the SN~1006 case by 
\citet{yamazaki2004b} may be possible,
although the filaments 
are very clumpy and we may have to consider the curvature effect 
suggested by \citet{berezhko2003}.
Together with the shock velocity of 562~km~s$^{-1}$ \citep{ghavamian2001}
and results of spatial and spectral analysis (see previous sections),
we estimated the magnetic field around the filaments is
$\sim$~4--12~$\mu$G \citep[see also][]{bamba2004}.
\citet{watanabe2003} reported on TeV gamma-ray observations
of the SW region of this remnant using the CANGAROO telescope.
According to our results,
inverse Compton TeV gamma-rays 
may be detected
with the current TeV $\gamma$-ray telescopes.
The estimated flux of inverse Compton emission is
$\sim6\times10^{-13}(B_{\rm d}/10\ \mu{\rm G})^{-2}$~ergs~s$^{-1}$cm$^{-2}$.

\subsection{Correlations between the age and spatial and spectral parameters}
\label{sec:evo}

We found thin and nonthermal filaments in four historical SNRs,
which may be acceleration sites of high-energy electrons.
Their scale widths and spectral parameters include a lot of information
about physical parameters inevitable
for understanding acceleration on the shock,
such as the gyro radius of electrons, the magnetic field configuration,
and so on,
as already suggested by \citet{bamba2003b} in the SN~1006 case.
Furthermore, a correlation between the age of SNRs
and the spatial and spectral parameters
might represent their time evolution in a single SNR,
although we should consider the characteristics of the individual source.
In this section,
we consider some correlation between the age and the physical parameters
as a rough discussion about the time evolution.

\begin{figure}[hbtp]
\epsscale{0.45}
\plotone{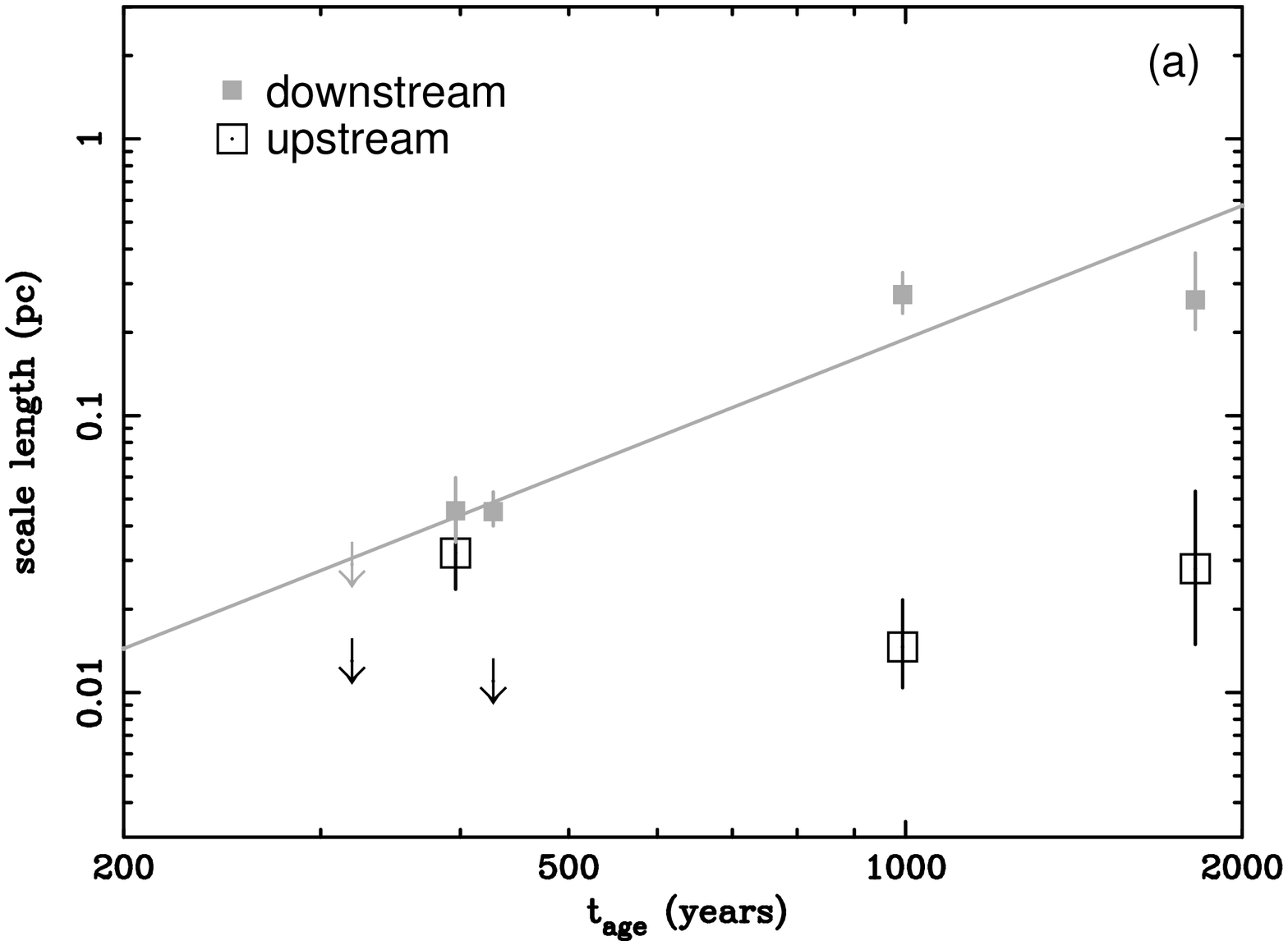}
\plotone{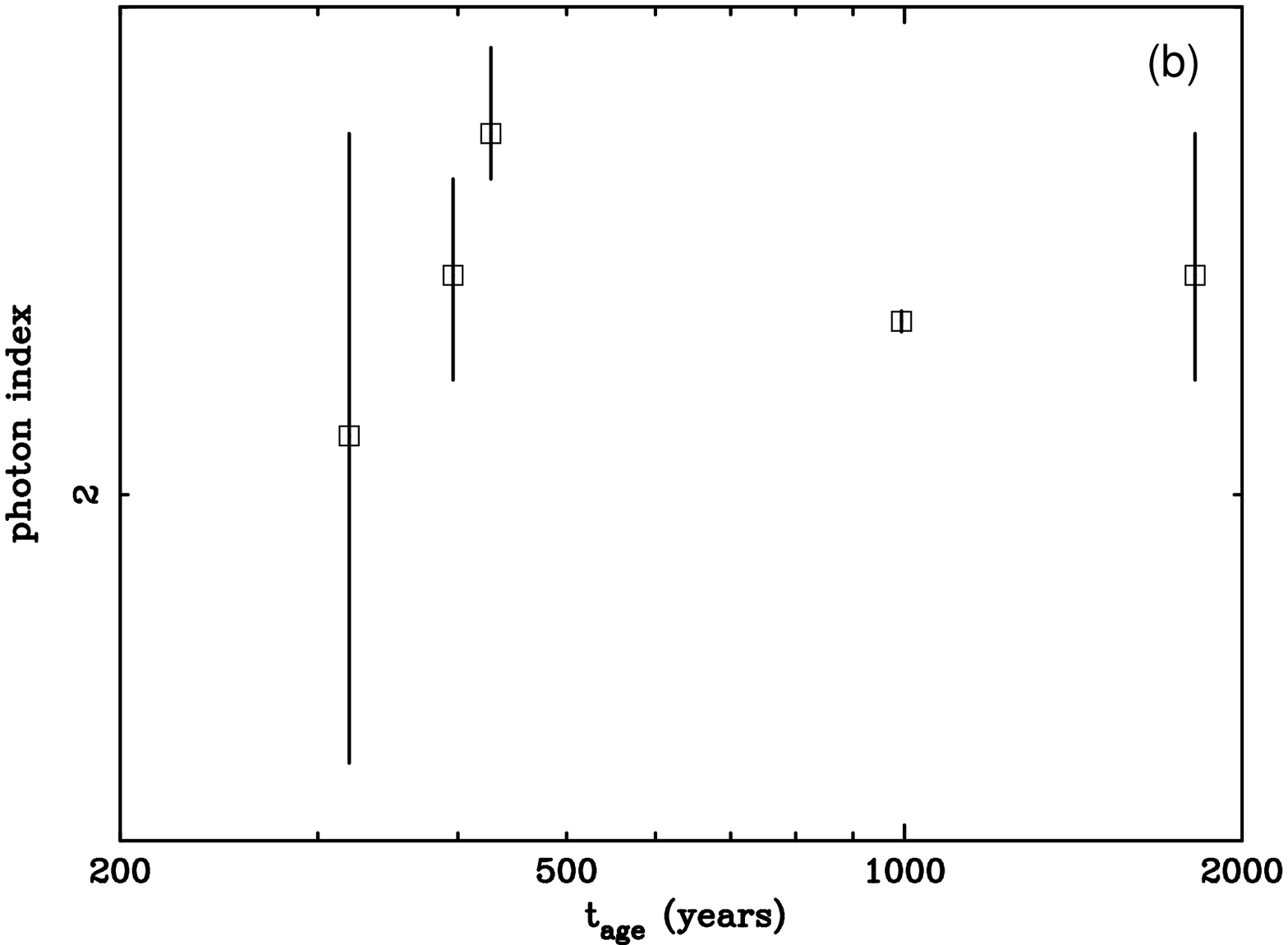}
\plotone{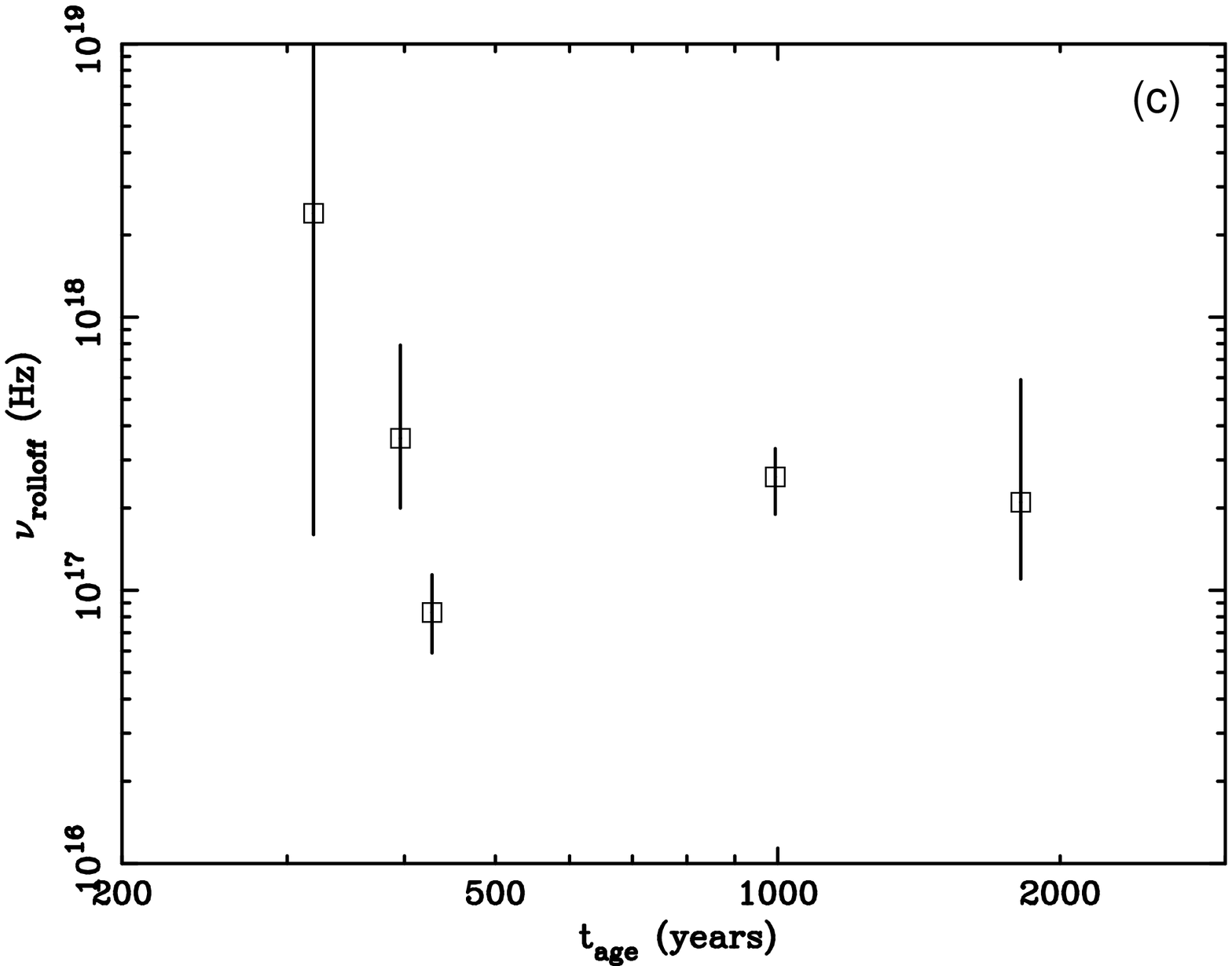}
\plotone{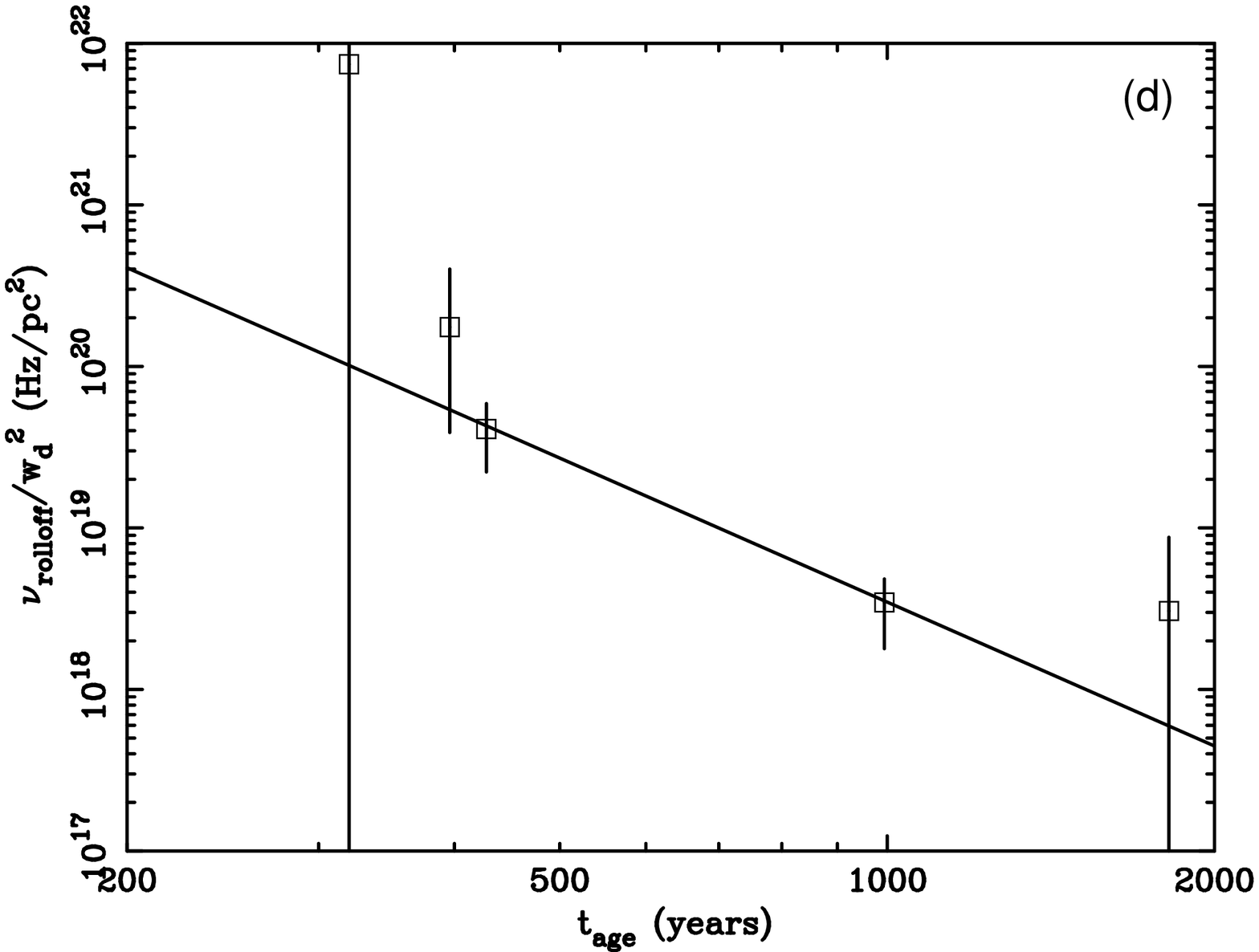}
\caption{Time evolutions of parameters,
(a) time vs. scale width,
(b) photon index of power-law fitting,
(c) $\nu_{\rm rolloff}$ of {\tt SRCUT} model fitting,
and (d) ${\cal B} \equiv \nu_{\rm rolloff}/w_d{}^2$.
Solid lines represent best-fit models for each plot (see text).}
\label{fig:evolution}
\end{figure}

Figure~\ref{fig:evolution}(a) represents
the average scale width
weighted by the flux of each filament
 ($w_u$ and $w_d$ in the unit of pc)
as a function of the age ($t_{age}$ in the unit of year)
for five historical SNRs;
we analyzed four and one was from the previous results of SN~1006
\citep{bamba2003b}.
The distances to the SNRs are fixed to previously estimated values
(see \S\ref{sec:individual}).
The panel shows that in all SNRs,
both $w_u$ and $w_d$ are quite smaller than the SNR radii.
Moreover, there is a positive correlation between $w_d$ and $t_{age}$;
the scale width becomes larger as the SNR becomes older.
We fitted them with a power-law function
as a tentative model,
and obtain the relations such as:
\begin{equation}
w_d = 3.0^{+0.7}_{-0.6}\times 10^{-6} t_{age}{}^{1.60_{-0.30}^{+0.01}}\ \ ,
\label{eq:wd}
\end{equation}
with the reduced $\chi^2$ of 9.68/3.
Hereinafter, the errors indicate 90\% confidence regions.
The index is larger than that for the shock width of thermal gas
derived from the Sedov solution (=2/5),
although we reached no definite conclusion
because the fittings were rejected statistically.

Figures~\ref{fig:evolution}(b) and (c) show the plots of 
photon indices in the power-law fittings vs. the age ($t_{age}$)(b)
and $\nu_{rolloff}$ (Hz) of the {\tt SRCUT} model vs. $t_{age}$ (c),
respectively.
The spectrum seems to grow softer
in the very young phase (less than 500~years),
whereas the spectral parameters remain constant at 1000--2000~years old.
Since the error is too large, especially in the young phase,
we gave up modeling the time evolution of the spectral parameters.

We searched for another parameter that has a clear correlation with time,
and found a tentative function, such as
\begin{equation}
{\cal B} = \nu_{rolloff}w_d{}^{-2}\ \ .
\end{equation}
Figure~\ref{fig:evolution}(d) represents
the relation between $t_{age}$ and ${\cal B}$.
It shows monotonous decay,
and a power-law fitting (${\cal B}=Ct_{age}{}^\alpha$)
was accepted statistically
with best-fit values of
\begin{eqnarray}
C &=& 2.6_{-1.4}^{+1.2}\times 10^{27}\ \, \\
\alpha &=& -2.96_{-0.06}^{+0.11}\ \ , \label{eq:f_index}
\end{eqnarray}
with a reduced $\chi^2$ of 2.03/3, respectively.
This result implies that
the function ${\cal B}$ might include some physical quantities
that evolve with time.
We checked the uncertainty of ${\cal B}$ due to the change of $s$
as already discussed in \S\ref{sec:origin},
and found that
the best-fit value does not change
although error regions become 10 times larger.
Therefore, the uncertainty of $s$ cannot change our results.

Here, we make a scenario in order to explain the time evolution of ${\cal B}$.
Our assumption is that the spatial profile found in our present analysis
reflects that of accelerated electrons \citep{bamba2003b,yamazaki2004b}.
The maximum energy of accelerated electrons, $E_{max}$, is determined by
the age of the SNR, or the synchrotron energy loss,
such as $t_{\rm acc}\sim{\rm min}\{t_{\rm age},t_{\rm loss}\}$,
where $t_{\rm acc}$ and $t_{\rm loss}$
are the acceleration and the synchrotron loss time scale,
respectively \citep{yamazaki2004b}.
Then, we simply obtain the scale width of
the accelerated electrons in the radial direction, $\Delta_d$, as
$\Delta_d\sim v_d \times{\rm min}\{t_{\rm age},t_{\rm loss}\}
\sim v_d t_{\rm acc}$,
where $v_d$ is the downstream fluid velocity.
The acceleration time $t_{\rm acc}$ is on the order of $K/v_s^2$,
where $v_s$ and $K$
are the shock velocity and
the diffusion coefficient, respectively.
The quantity $K$ is assumed to
be proportional to the gyro radius of the accelerated electrons,
$r_g=E_{max}/eB_d$, where $B_d$ is
the downstream magnetic field  \citep{drury1983,yamazaki2004b}.
As a result, we find
$\Delta_d\sim K/v_s\propto B_d{}^{-1}E_{max}v_s{}^{-1}$.
Because of the projection (curvature) effect,
observed apparent scale width in the downstream region, $w_d$,
differs from $\Delta_d$ \citep[e.g.,][]{berezhko2004}.
The ratio $w_d/\Delta_d$ depends on uncertainty of
radial profile of accelerated electron distribution downstream.
It is found that for sufficiently small
$\Delta_d$ compared with the local curvature radius,
the ratio $w_d/\Delta_d$ is about 7, 5, or 1
when we assume the exponential, Gaussian, or the top-hat form of the
downstream radial profile, respectively.
Hence it may be assumed that the ratio $w_d/\Delta_d(\sim3)$ remains
unchanged within a factor of 3, so we find
$w_d\propto\Delta_d\propto B_d{}^{-1}E_{max}v_s{}^{-1}$.
The value of $\nu_{rolloff}$ is proportional to $B_dE_{max}{}^2$
\citep{reynolds1999}.
Therefore, the quantity ${\cal B}$ is proportional to $B_d{}^3v_s{}^2$.
The time evolution of ${\cal B}$ may represent
that of $B_d{}^3v_s{}^2$.

Let us consider the time evolution of the magnetic field and the shock velocity
as $B_d \propto t_{age}{}^{-\ell}$ and $v_s \propto t_{age}{}^{-m}$.
The index, $m$, changes from 0 to $3/5$
as SNRs evolve from the free-expansion phase to the Sedov phase.
Together with the observational fact of eq.(\ref{eq:f_index}),
the index $\ell$ takes the value of 0.57--1.02.
Our result suggests that
the magnetic field decreases as the SNR ages.
This is the first observational implication
of the time evolution of the magnetic field in the SNR.

The most interesting case is $\ell=0.6$,
which is allowed when SNRs are in the Sedov phase ($m=0.6$).
In this case, since the density of the thermal plasma is almost constant,
the energy density of the magnetic field ($u_B \propto B_d{}^2$),
and the kinetic and thermal energy densities of the shock ($u_{th} \propto$
shock temperature
and $u_{kin} \propto v_s{}^2$), 
evolves according to the same time dependency (index = $-1.2$).
The time evolution of the energy density of accelerated protons ($u_{p}$)
may also be discussed
from the relations by \citet{lucek2000} as
\begin{eqnarray}
u_{p} &=& \frac{v_s}{v_A} u_B \propto B_dv_s
\propto t_{age}{}^{-1.2}\ \ ,
\end{eqnarray}
where $v_A \equiv B/\sqrt{4\pi\rho}$ and $\rho$ are the Alfv{\'e}n velocity
and the fluid density,
suggesting that
accelerated protons in an SNR evolve with the same energy density evolutions
as that of the other energy carriers.
Since \citet{bamba2003b} indicated that 
these energy densities are roughly in equipartition with each other,
our result implies that
they evolve while maintaining equipartition.
\citet{bell2001} suggested theoretically that
the magnetic field on the shock evolves 
according to the same time dependency as
the shock velocity,
which is consistent with our result.

There remain many unsolved problems that we should consider.
The value of $w_d$ depends on many parameters that we have ignored,
such as the curvature effect,
the angle between the magnetic field and the shock normal,
the degree of turbulence, and so on.
Solving these problems is beyond our work,
which is going to be considered in \citet{yamazaki2004a}.

\section{Summary}
\label{sec:sum}

We have conducted systematic spectral and spatial analysis of
filamentary structures
in historical SNRs for the first time.
A summary of our results is as follows:

\begin{enumerate}
\item
We discovered very thin filaments on the outer edges of four historical SNRs:
Cas~A, Kepler, Tycho, and RCW~86.
The scale widths were measured for the first time;
they are below or the same as the PSF size of {\it Chandra} and 0.01--0.4~pc
on the upstream and downstream sides of the emission peak, respectively.
\item
Although
both thermal and nonthermal spectral models can reproduce
the spectra of the filaments,
the former is unlikely since
it requires an unrealistic high temperature and low abundances.
The photon indices ($\sim$2--3) and $\nu_{rolloff}$
($\sim 10^{16}$--$10^{18}$~Hz) of nonthermal models suggest
the synchrotron X-ray emission from electrons accelerated via DSA,
in a similar way to the SN~1006 case.
\item
The value of $w_d$ (see eq.(\ref{eq:wd})) becomes
larger as the SNR ages.
\item
The spectra of filaments might become softer as the SNR ages
in the phase of $t_{age} \la 500$~years,
whereas it does not change when $1000 \la t_{age} \la 2000$~years,
although there is no definite conclusion because of a lack of statistics.
\item
The time evolution of a tentative function
${\cal B} = \nu_{rolloff}/w_d{}^2$
was analyzed,
and it has a clear negative correlation
with $t_{age}$ with the index of $\sim -3$.
This may imply that
the downstream magnetic field decreases 
with the index ($\ell$) of 0.6--1.0.
When $\ell=0.6$, 
energy densities of magnetic field, shock (thermal and kinetic),
and cosmic rays evolve
while keeping equipartition with each other.
\end{enumerate}

\acknowledgements

Our particular thanks are due to
M. Hoshino, S. Inutsuka, K. Makishima, and F. Takahara,
for their fruitful discussions and comments.
We also thank an anonymous referee
for helpful comments.
R.Y. is supported by JSPS Research Fellowship for Young Scientists.

\onecolumn

\begin{deluxetable}{p{7pc}cccccc}
\tabletypesize{\scriptsize}
\tablecaption{Observation Log
\label{tab:obs_log}}
\tablewidth{0pt}
\tablecolumns{2}
\tablehead{
\colhead{Target} & \colhead{Type} & \colhead{ObsID} & \colhead{R.A.} &
\colhead{Dec.} & \colhead{Date} & \colhead{Exposure (ks)} 
}
\startdata
Cas A\dotfill & Ib & 00114 & 23\fh23\fm40\fs2 & 58\fd47\fm34\fs1 & 2000 Jan. 30--31 & 50 \\
Kepler\dotfill & ? & 00116 & 17\fh30\fm39\fs6 & -21\fd30\fm32\fs5 & 2000 Jun.\ 30 -- Jul.\ 1 & 49 \\
Tycho\dotfill & Ia & 00115 & 10\fh25\fm07\fs0 & 64\fd09\fm44\fs7 & 2000 Sep.\ 20--21 & 49 \\ 
SN~1006\dotfill & Ia & 00732 & 15\fh03\fm51\fs6 & -41\fd51\fm18\fs8 & 2000 Jul.\ 10--11 & 68 \\ 
SW shell of RCW~86\dotfill & II? & 01993 & 14\fh40\fm46\fs6 & -62\fd39\fm43\fs9 & 2001 Feb. 1--2 & 92 
\enddata
\end{deluxetable}

\begin{deluxetable}{p{4pc}cccc}
\tabletypesize{\scriptsize}
\tablecaption{Best-fit scale width of filaments\tablenotemark{a}
\label{tab:fila_image}}
\tablewidth{0pt}
\tablecolumns{2}
\tablehead{
\colhead {No.} & \colhead{$A$} & \colhead{$w_u$} & \colhead{$w_d$} & \colhead{reduced $\chi^2$} \\
 & [counts arcsec$^{-1}$] & [arcsec] & [arcsec] & [$\chi^2$/d.o.f.]
}
\startdata
\multicolumn{5}{c}
{Cas A} \\ \hline
1\dotfill & 59 (46--72) & ($<$0.93) & 1.27 (0.96--1.81) & 99.7/61 \\
2\dotfill & 34 (24--49) & ($<$0.80) & ($<$1.59) & 27.3/26 \\
\hline
\multicolumn{5}{c}
{Kepler} \\ \hline
1\dotfill & 33 (26--41) & 1.17 (0.87--1.59) & 0.93 ($<$1.41) & 19.3/21 \\
2\dotfill & 29 (25--34) & 1.59 (1.17--2.19) & 3.09 (3.46--3.87) & 70.1/47 \\
\hline
\multicolumn{5}{c}
{Tycho} \\ \hline
1\dotfill & 129 (119--139) & 1.18 (1.01--1.32) & 5.36 (4.47--6.12) & 82.7/55 \\
2\dotfill & 62 (48--70) & ($<$0.80) & 1.70 (1.32--3.15) & 42.3/32 \\
3\dotfill & 80 (74--83) & ($<$0.80) & 2.38 (2.20--2.54) & 23.4/37 \\
4\dotfill & 150 (146--152) & 0.86 (0.80--0.93) & 5.53 (5.00--6.14) & 48.9/46 \\
5\dotfill & 63 (57--71) & 1.03 (0.90--1.35) & 2.47 (1.93--3.15) & 17.9/21 \\
\hline
\multicolumn{5}{c}
{RCW~86} \\ \hline
1\dotfill & 28 (25--30) & 2.39 (1.48--3.34) & 20.1 (17.3--23.8) & 113.4/95 \\
2\dotfill & 18 (14--23) & 1.56 (0.49--4.79) & 18.2 (11.8--35.6) & 46.8/31 
\enddata
\tablenotetext{a}{Parentheses indicate single parameter 90\% confidence regions.}
\end{deluxetable}

\begin{deluxetable}{p{2pc}cccccccccccccc}
\tabletypesize{\scriptsize}
\tablecaption{Best-fit parameters of spectral fittings of
filaments%
\tablenotemark{a}
\label{tab:fila_spec}}
\rotate
\tablewidth{0pt}
\tablecolumns{2}
\tablehead{
 & \multicolumn{6}{c}{\tt NEI} & \multicolumn{4}{c}{\tt power-law} & \multicolumn{4}{c}{\tt SRCUT\tablenotemark{b}} \\
\cline{2-7} \cline{8-11} \cline{12-15}
\colhead{No.} &
\colhead{$kT$ [keV]} & \colhead{$Z$\tablenotemark{c}} & \colhead{$n_et$\tablenotemark{d}} & \colhead{$E.M.$\tablenotemark{e}} & \colhead{$N_{\rm H}$\tablenotemark{f}} & \colhead{$\chi^2$/d.o.f.} &
\colhead{$\Gamma$} & \colhead{$N_{\rm H}$\tablenotemark{f}} & \colhead{Flux\tablenotemark{g}} & \colhead{$\chi^2$/d.o.f.} &
\colhead{$\nu_{rolloff}$\tablenotemark{h}} & \colhead{$N_{\rm H}$\tablenotemark{f}} & \colhead{$\Sigma_{\rm 1GHz}$\tablenotemark{i}} & \colhead{$\chi^2$/d.o.f.}
}
\startdata
\multicolumn{15}{c}{Cas A}
\\ \hline
1\dotfill &
5.4 & 0.28 & \nodata & 3.6 & 27 & 28.0/31 &
2.3  & 27 & 1.9 & 29.2/33 & 
9.3 & 26 & 6.3 & 29.2/33 \\
 & ($>$2.1) & ($<$1.4) & \nodata\tablenotemark{j} & (2.0--8.1) & (15--41) & &
(2.0--3.5) & (20--47) & & &
($>4.5$) & (20--33) & (5.1--7.6) &  
q\\
2\dotfill &
5.7 & \nodata & 0.030 & 0.15 & 17 & 19.1/27 &
2.2 & 20 & 1.1 & 22.1/29 &
28 & 20 & 2.1 & 22.1/29 \\
 & (2.7--34) & ($>$0.19) & (0.011--0.096) & (0.073--3.5) & (11--25) & &
(1.9--2.8) & (13--31) & & & 
$(>1.3)$ & (16--24) & (1.7--2.4) & 
\\
total\dotfill & 
5.4 & 0.26 & 3.8 & 5.6 & 25 & 56.4/42 & 
2.1 & 23 & 3.1 & 57.6/44 & 
24 & 22 & 6.3 & 57.6/44 \\
 & (3.0--13) & (0.019--0.73) & (1.6--7.7) & (4.1--9.3) & (17--35) & & 
(1.6--2.7) & (15--34) & & & 
($>$1.8) & (19--27) & (5.4--7.2) &
\\
\hline
\multicolumn{15}{c}{Kepler}\\
\hline
1\dotfill &
3.5 & \nodata & 8.2 & 5.5 & 4.3 & 43.2/52 & 
2.3 & 5.8 & 1.7 & 42.2/54 & 
5.5 & 5.3 & 0.90 & 42.8/54 \\
 & (2.7--4.9) & ($<0.067$) & (3.4--28) & (4.5--6.8) & (3.4--5.6) & & 
(2.1--2.5) & (4.8--7.0) & & & 
(2.1--17) & (4.7--6.0) & (0.83--0.98) \\
2\dotfill & 
3.0 & 0.13 & \nodata & 7.9 & 3.7 & 25.7/30 & 
2.4 & 5.5 & 2.2 & 24.9/32 &
4.0 & 4.9 & 1.4 & 25.0/32 \\
& (2.5--3.8) & ($<0.40$) & \nodata\tablenotemark{j} & (6.5--9.7) & (3.0--4.5) & & 
(2.2--2.6) & (4.5--6.7) & & & 
(1.8--9.8) & (4.3--5.5) & (1.3--1.5) & \\
total\dotfill & 
3.2 & \nodata & 11 & 14 & 4.0 & 47.6/53 & 
2.4 & 5.7 & 4.1 & 44.9/55 & 
3.6 & 5.2 & 2.8 & 45.5/55 \\
 & (2.6--3.9) & ($<$0.34) & ($>$6.9) & (12--17) & (3.4--4.7) & & 
(2.2--2.6) & (4.9--6.6) & & & 
(2.0--7.9) & (4.8--5.7) & (2.6--3.0) & \\
\hline
\multicolumn{15}{c}{Tycho}\\
\hline
1\dotfill &
1.9 & 0.22 & $3.4\times 10^3$ & 4.7 & 6.5 & 102.3/68 & 
3.0 & 9.1 & 3.4 & 104.8/70 & 
0.40 & 8.1 & 0.70 & 106.0/70 \\
& (1.6--2.4) & (0.062--0.50) & ($>$69) & (3.5--6.0) & (5.4--7.5) & & 
(2.8--3.3) & (7.9--11) & & & 
(0.25--0.70) & (7.4--8.8) & (0.64--0.77) & \\
2\dotfill &
1.0 & \nodata & \nodata & 2.5 & 12 & 9.40/11 & 
3.8 & 16 & 0.49 & 9.05/13 & 
0.11 & 14 & 1.4 & 9.15/13 \\
 & (0.58--6.9) & ($<$1.6) & \nodata\tablenotemark{j} & (0.27--9.2) & (6.6--20) & & 
(2.5--5.6) & (8.6--27) & & & 
(0.020--1.3) & (11--19) & (1.0--2.0) & \\
3\dotfill & 
2.1 & 0.12 & \nodata & 1.7 & 11 & 40.2/41 & 
2.5 & 7.6 & 1.1 & 43.5/43 & 
1.3 & 6.9 & 0.051 & 43.2/43 \\
& (1.6--2.9) & ($<$0.37) & \nodata\tablenotemark{j} & (0.85--2.4) & (5.7--15) & &
(2.2--2.9) & (6.0--9.8) & & & 
(0.52--4.5) & (5.9--8.2) & (0.045--0.059) & \\
4\dotfill & 
4.3 & \nodata & \nodata & 2.9 & 5.1 & 39.5/32 &
2.1 & 6.8 & 4.4 & 40.9/34 &
6.5 & 6.2 & 0.052 & 40.6/34 \\
 & (3.0--7.0) & ($<$0.38) & \nodata\tablenotemark{j} & (2.3--3.9) & (3.4--10) & & 
(2.0--2.4) & (4.6--9.6) & & & 
(1.8--46) & (5.0--7.7) & (0.046--0.059) & \\
5\dotfill & 
1.8 & 0.05 & \nodata & 2.0 & 9.8 & 51.9/47 & 
2.9 & 9.4 & 1.1 & 55.5/49 & 
0.64 & 8.2 & 0.12 & 54.4/49 \\
 & (1.4--2.6) & ($<$0.20) & \nodata\tablenotemark{j} & (1.1--3.2) & (5.1--14) & & 
(2.5--3.3) & (6.9--13) & & & 
(0.26--1.7) & (6.7--10) & (0.10--0.14) & \\
total\dotfill &
2.3 & 0.07 & \nodata & 12 & 5.8 & 182.0/134 & 
2.7 & 8.0 & 11 & 187.7/136 & 
0.83 & 7.2 & 0.78 & 185.2/136 \\
 & (2.0--2.6) & ($<$0.17) & \nodata\tablenotemark{j} & (11--14) & (5.2--6.4) & &
(2.6--2.9) & (7.3--9.0) & & & 
(0.59--1.2) & (6.7--7.7) & (0.73--0.83) & \\
\hline
\multicolumn{15}{c}{RCW~86}\\
\hline
1\dotfill & 
2.5 & 0.032 & \nodata & 2.1 & 5.7 & 67.2/58 & 
2.5 & 5.4 & 1.4 & 68.9/60 &
2.0 & 4.8 & 0.19 & 68.8/60 \\
 & (1.8--3.4) & ($<$0.12) & \nodata\tablenotemark{j} & (1.6--2.8) & (3.3--8.3) & & 
(2.2--2.8) & (4.3--6.9) & & & 
(0.77--5.9) & (4.2--5.7) & (0.17--0.22) & \\
2\dotfill & 
1.4 & 0.27 & 0.35 & 0.82 & 7.0 & 18.7/25 &
2.5 & 2.7 & 0.32 & 25.7/27 &
1.3 & 2.3 & 0.051 & 25.0/27 \\
& (1.0--2.2) & (0.019--1.0) & (0.20--0.52) & (0.49--1.4) & (3.2--9.4) & & 
(2.1--3.2) & (1.1--4.1) & & & 
(0.29--12) & (1.5--3.5) & (0.039--0.064) & \\
total\dotfill &
2.1 & 0.11 & 0.43 & 3.3 & 7.0 & 75.3/85 & 
2.4 & 4.7 & 1.8 & 82.2/87 & 
2.1 & 4.3 & 0.22 & 80.8/87 \\
 & (1.6--2.6) & ($<$0.25) & ($<$0.80) & (2.1--4.3) & (3.1--9.1) & & 
(2.2--2.7) & (4.2--6.0) & & & 
(1.0--5.9) & (3.7--5.0) & (0.20--0.25) & 
\enddata
\tablenotetext{a}
{Parentheses indicate single parameter 90\% confidence regions.}
\tablenotetext{b}
{Photon index at 1~GHz is assumed according to the previous radio observations
(see text).}
\tablenotetext{c}
{Abundance ratio relative to the solar value \citep{anders1989}.}
\tablenotetext{d}
{Ionization time-scale
in the unit of  $10^{10}$~s~cm$^{-3}$, where
$n_e$ and $t$ are the electron density and age of the plasma.}
\tablenotetext{e}
{Emission measure
in the unit of 10$^{55}$cm$^{-3}$,
with the assumed distance to SNRs according to the previous observations
(see text).}
\tablenotetext{f}
{Absorption column in the unit of $10^{21}$cm$^{-2}$,
calculated using the cross sections by \citet{morrison1983}
with the solar abundances \citep{anders1989}.}
\tablenotetext{g}
{Observed flux in the 0.5--10.0~keV band in the unit of $10^{-13}$~ergs~cm$^{-2}$s$^{-1}$.}
\tablenotetext{h}
{Roll-off frequency in the unit of $10^{17}$~Hz.}
\tablenotetext{i}
{Flux density at 1~GHz in the unit of 0.1~Jy.}
\tablenotetext{j}
{Not determined.}
\end{deluxetable}


\begin{thebibliography}{}
\bibitem[Allen et al.(1997)]{allen1997}
Allen, G.~E.~et al.\ 1997, \apjl, 487, L97 
\bibitem[Anders \& Grevesse(1989)]{anders1989}
Anders, E., \& Grevesse, N. 1989, \gca, 53, 197
\bibitem[Bamba(2004)]{bamba2004}
Bamba, A.\ 2004, PhD thesis (Kyoto University)
\bibitem[Bamba et al.(2000)Bamba, Tomida, \& Koyama]{bamba2000}
Bamba, A., Tomida, H., \& Koyama, K. 2000, \pasj, 52, 1157
\bibitem[Bamba et al.(2001)]{bamba2001} 
Bamba, A., Ueno, M., Koyama, K., \& Yamauchi, S.\ 2001, \pasj, 53, L21 
\bibitem[Bamba et al.(2003a)]{bamba2003a} 
Bamba, A., Ueno, M., Koyama, K., \& Yamauchi, S.\ 2003a, \apj, 589, 253 
\bibitem[Bamba et al.(2003b)]{bamba2003b} 
Bamba, A., Yamazaki, R., Ueno, M., \& Koyama, K.\ 2003b, \apj, 589, 827 
\bibitem[Bell(1978)]{bell1978}
Bell, A.~R.\ 1978, \mnras, 182, 443
\bibitem[Bell \& Lucek(2001)]{bell2001}
Bell, A.~R.~\& Lucek, S.~G.\ 2001, \mnras, 321, 433
\bibitem[Berezhko et al.(2003)Berezhko, Ksenofontov, \& V{\" o}lk]
{berezhko2003}
Berezhko, E.~G., Ksenofontov, L.~T., \& V{\" o}lk, H.~J.\ 
2003, \aap, 412, L11
\bibitem[Berezhko \& V{\" o}lk(2004)]{berezhko2004}
Berezhko, E.~G., \& V{\" o}lk, H.~J.\ 2004, \aap, accepted (astroph/0404203)
\bibitem[Blandford \& Eichler(1987)]{blandford1987}
Blandford, R.~D., \& Eichler, D.\ 1987, \physrep, 154,1
\bibitem[Blandford \& Ostriker(1978)]{blandford1978}
Blandford, R.~D., \& Ostriker, J.~P.\ 1978, \apj, 221, L29
\bibitem[Borkowski et al.(2001a)]{borkowski2001a}
Borkowski, K.J., Lyerly, W.J., \& Reynolds, S.P. 2001a, \apj, 548, 820
\bibitem[Borkowski et al.(2001b)]{borkowski2001b}
Borkowski, K.~J., Rho, J., Reynolds, S.~P. \& Dyer, K.K. 2001b, \apj, 550, 334
\bibitem[Cassam-Chena{\" i} et al.(2003)]{cassamchenai2004}
Cassam-Chena{\" i}, G., Decourchelle, A., Ballet, J.,
Hwang, U., Hughes, J.~P., \& Petre, R.\ 2004, \aap, 414, 545
\bibitem[Caswell et al.(1975)Caswell, Clark, \& Crawford]{caswell1975}
Caswell, J.~L., Clark, D.~H., \& Crawford, D.~F.\ 1975, Australian Journal of 
Physics Astrophysical Supplement, 39 
\bibitem[Chin \& Huang(1994)]{chin1994}
Chin, Y.~N.~\& Huang, Y.~L.\ 1994, \nat, 371, 398 
\bibitem[Clark \& Stephenson(1977)]{clark1977}
Clark, D.~H.~\& Stephenson, F.~R.\ 1977, Oxford; (New York: Pergamon Press, 
1977.~1st ed.)
\bibitem[Decourchelle et al.(2001)]{decourchelle2001}
Decourchelle, A.~et al.\ 2001, \aap, 365, L218 
\bibitem[DeLaney et al.(2002)]{delaney2002} 
DeLaney, T., Koralesky, B., Rudnick, L., \& Dickel, J.~R.
2002, \apj, 580, 914 
\bibitem[Drury(1983)]{drury1983}
Drury, L.O'C.\ 1983, Rep. Prog. Phys., 46, 973
\bibitem[Favata et al.(1997)]{favata1997}
Favata, F.~et al.\ 1997, \aap, 324, L49 
\bibitem[Fan You(432)]{fanyou432}
Fan You\ 432, ``Houhanshu'', 6, 3260
\bibitem[Fink et al.(1994)]{fink1994}
Fink, H.~H., Asaoka, I., Brinkmann, W., Kawai, N., \& Koyama, K.\ 1994,
\bibitem[Flamsteed(1725)]{flamsteed1725}
Flamsteed, J.\ 1725, "Historia Coelestia star catalog"
\bibitem[Garmire et al.(2000)]{garmire2000}
Garmire, G., Feigelson, E.~D., Broos, P., Hillenbrand, L.~A.,
Pravdo, S.~H., Townsley, L.,\&  Tsuboi, Y.\ 2000, \aj, 120, 1426
\bibitem[Ghavamian et al.(2001)]{ghavamian2001}
Ghavamian, P., Raymond, J., Smith, R.~C., \& Hartigan, P.\ 2001, \apj. 548, 995
\bibitem[Green(2004)]{green2004}
Green, D.~A.\ 2004,
A Catalogue of Galactic Supernova Remnants (2004 January version), 
(Cambridge, UK, Mullard Radio Astronomy Observatory)
available on the WWW at http://www.mrao.cam.ac.uk/surveys/snrs/
\bibitem[Hess(1912)]{hess1912}
Hess, V.~F.\ 1912, Phys. Zeits., 13, 1084
\bibitem[Hester(1987)]{hester1987}
Hester, J.~J.\ 1987, \apj, 314, 187
\bibitem[Hwang et al.(2002)]{hwang2002} 
Hwang, U., Decourchelle, A., Holt, S.~S., \& Petre, R.\ 2002, \apj, 581, 1101 
\bibitem[Jones \& Ellison(1991)]{jones1991}
Jones, F.C., \& Ellison, D.C.\ 1991, Space Science Rev., 58, 259
\bibitem[Kaastra et al.(1992)]{kaastra1992} 
Kaastra, J.~S., Asaoka, I., Koyama, K., \& Yamauchi, S.\ 1992, \aap, 264, 654 
\bibitem[Kamper \& van den Bergh(1978)]{kamper1978}
Kamper, K.~W.~\& van den Bergh, S.\ 1978, \apj, 224, 851 
\bibitem[Katz-Stone et al.(2000)]{katzstone2000}
Katz-Stone, D.~M., Kassim, N.~E., 
Lazio, T.~J.~W., \& O'Donnell, R.\ 2000, \apj, 529, 453
\bibitem[Lucek \& Bell(2000)]{lucek2000}
Lucek, S.~G.~\& Bell, A.~R.\ 2000, \mnras, 314, 65 
\bibitem[Kepler(1606)]{kepler1606}
Kepler, J.\ 1606. ``De stella nova''
\bibitem[Koyama et al.(1997)]{koyama1997}
Koyama, K., Kinugasa, K., Matsuzaki, K., Nishiuchi, M., Sugizaki, M.,
Torii, K., Yamauchi, S., \& Aschenbach, B.\ 1997, \pasj, 49, L7 
\bibitem[Koyama et al.(1995)]{koyama1995}
Koyama, K., Petre, R., Gotthelf, E.V., Hwang, U., Matsura, M., 
Ozaki, M., \& Holt S.~S.\ 1995, \nat, 378, 255
\bibitem[Long et al.(2003)]{long2003}
Long, K.~S., Reynolds, S.~P., Raymond, J.~C., Winkler, P.~F.,
Dyer, K.~K., \& Petre, R.\ 2003, \apj, 586, 1162 
\bibitem[Malkov \& Drury(2001)]{malkov2001}
Malkov, E., \& Drury, L.O'C.\ 2001, Rep.\ Prog.\ Phys., 64, 429
\bibitem[Morrison \& McCammon(1983)]{morrison1983}
Morrison, R., \& McCammon, D.\ 1983, \apj, 270, 119
\bibitem[Petre et al.(1999)Petre, Allen, \& Hwang]{petre1999}
Petre, R., Allen, G.~E., Hwang, U.\ 1999,
Astronomische Nachrichten 320, 199
\bibitem[Pravdo \& Smith(1979)]{pravdo1979}
Pravdo, S.~H., \& Smith, B.~W.\ 1979, \apj, 234, L195
\bibitem[Reed et al.(1995)]{reed1995}
Reed, J.~E., Hester, J.~J., Fabian, A.~C., \& Winkler, P.~F.\ 1995, \apj, 440, 
706 
\bibitem[Reynolds(1998)]{reynolds1998}
Reynolds, S.~P.\ 1998, \apj, 493, 375
\bibitem[Reynolds \& Keohane(1999)]{reynolds1999}
Reynolds, S.P., \& Keohane, J.W.\ 1999, \apj, 525, 368
\bibitem[Reynoso \& Goss(1999)]{reynoso1999}
Reynoso, E.~M.~\& Goss, W.~M.\ 1999, \aj, 118, 926 
\bibitem[Rho et al.(2002)]{rho2002}
Rho, J., Dyer, K.~K., Borkowski, K.~J., \& Reynolds, S.~P.\ 2002, \apj, 581, 
1116 
\bibitem[Rosado et al.(1996)]{rosado1996}
Rosado, M., Ambrocio-Cruz, P., Le Coarer, E., \& Marcelin, M.\ 
1996, \aap, 315, 243 
\bibitem[Slane et al.(1999)]{slane1999}
Slane, P., Gaensler, B.~M., Dame, T.~M., Hughes, J.~P., Plucinsky, P.~P.,
\& Green, A.\ 1999, \apj, 525, 357 
\bibitem[Slane et al.(2001)]{slane2001}
Slane, P., Hughes, J.~P., Edgar, R.~J., Plucinsky, P.~P., Miyata, E.,
Tsunemi, H., \& Aschenbach, B.\ 2001, \apj, 548, 814 
\bibitem[The et al.(1996)]{the1996}
The, L.-S., Leising, M.~D., Kurfess, J.~D., Johnson, W.~N.,
Hartmann, D.~H., Gehrels, N., Grove, J.~E., 
\& Purcell, W.~R.\ 1996, \aaps, 120, 357 
\bibitem[Tycho(1573)]{tycho1573}
Tycho, B.\ 1573, ``De Stella Nova'' (Copenhagen: Laurenentius)
\bibitem[Ueno et al.(2003)]{ueno2003}
Ueno, M., Bamba, A., Koyama, K., \& Ebisawa, K.\ 2003, \apj, 588, 338 
\bibitem[Vink \& Laming(2003)]{vink2003}
Vink, J.~\& Laming, J.~M.\ 2003, \apj, 584, 758 
\bibitem[Watanabe et al.(2003)]{watanabe2003}
Watanabe, S.\ et al.\ 2003,
proceedings of the 28th Internal Cosmic Ray Conference
\bibitem[Weisskopf et al.(2002)]{weisskopf2002}
Weisskopf, M.~C.,
Brinkman, B., Canizares, C., Garmire, G., Murray, S., \& Van Speybroeck,
L.~P.\ 2002, \pasp, 114, 1
\bibitem[Winkler et al.(2003)Winkler, Gupta, \& Long]{winkler2003}
Winkler, P.~F., Gupta, G., \& Long, K.~S.\ 2003, \apj, 585, 324 
\bibitem[Yamaguchi et al.(2004)]{yamaguchi2004}
Yamaguchi, H., Ueno, M., Bamba, A., \& Koyama, K.\ 2004, accepted by \pasj,
(astro-ph/040017)
\bibitem[Yamazaki et al.(2004a)]{yamazaki2004a}
Yamazaki, R., Bamba, A., Takahara, F., Yoshida, T., \& Terasawa, T., \ 2004a,
submitted to \apjl
\bibitem[Yamazaki et al.(2004b)]{yamazaki2004b}
Yamazaki, R., Yoshida, T., Terasawa, T., Bamba, A., \& Koyama, K.\ 2004b,
\aap, 416, 595
\end{thebibliography}
\end{document}